\documentclass[%
 aip,
 pof,
 amsmath,amssymb,
 preprint,
 numerical
]{revtex4-1}
\usepackage{amsmath}
\usepackage{epsfig}
\usepackage{graphicx}
\usepackage{dcolumn}
\usepackage{bm}

\newcommand{\ud}{\mathrm{d}}
\newcommand{\uF}{\mathrm{F}}
\newcommand{\ui}{\mathrm{i}}
\newcommand{\ue}{\mathrm{e}}
\newcommand{\uRe}{\mathrm{Re}}

\begin{document}

\title{A theoretical and numerical study of gravity driven coating flow on cylinder and sphere:
two-dimensional and axisymmetric capillary waves}

\author{Shuo Hou}
 \email{hs180934@163.com.}

\affiliation{
School of Aeronautic Science and Engineering, Beihang University, Beijing, China
}

\begin{abstract}
The theoretical and numerical models for gravity driven coating flow on upper cylinder and sphere are formulated. Using a perturbation method, the governing equations which depend on one Bond number $Bo$ are derived for a liquid film flow down the outside of a horizontal cylinder and a sphere. They can be simplified to one-dimensional form due to the symmetries. The general structure of the two-dimensional and axisymmetric capillary waves under high $Bo$ condition is focused on. An asymptotic theory is used to solve for the free-surface profiles in the outer and inner region, respectively. Even though the evolution in the outer region is essentially different, there are inherent similarities in the inner region because the capillary ridges are proved to degenerate into the one-parameter family in high $Bo$ limit. Using appropriate numerical techniques, some parametric studies are performed on the profiles of the capillary waves. The spreading of the front recorded from the outer solutions quantitatively accords with the scaling laws in a top region at early time, but deviates obviously at later time. The comparison between the composite solutions constructed using the asymptotic theory and direct solutions calculated from the evolution equations is highlighted via both the global and local features. Agreement between the composite solutions and the direct solutions is good for high $Bo$ cases. This asymptotic behavior is common on both of cylindrical and spherical surfaces and not affected by the partial wetting process.
\end{abstract}

\maketitle

\section{\label{sec:intro}Introduction}

In physicochemical hydrodynamics, the coating flow is a fluid flow in which a large solid surface area is covered with one or more thin liquid layers \citep{Probstein1994}. Examples of scientific and technical importance range from small scales \citep{Stone2004} (coating processes in manufacturing) to large scales \citep{Griffiths2000} (lava flow on volcanoes). The fluid flow is always coupled with some surface physicochemical processes such as wetting and spreading \citep{Gennes1985}. Even though the aim is often to study the way in which an extra quantity of liquid moves on an already wet wall, many of the coating flow problems contain the evolution of a moving contact line. The macroscopic hydrodynamic model breaks down at the contact line as the traditional no-slip condition causes a stress singularity. The no-slip condition can be relaxed by allowing slip in the vicinity of the contact line \citep{Dussan1976,Greenspan1978} or introducing a thin precursor film \citep{Tanner1979} on the solid substrate. These two approaches for dynamic contact line give rise to many coating flow researches in last two decades \citep{Craster2009}. Most of the researches focused on gravity \citep{Kondic2003,Lister2010} or thermally \citep{Kataoka1997} driven flow of a thin liquid film spreading on a solid surface. The region near the moving contact line is dominated by the surface tension effect. A capillary ridge is typically observed in this region, and the spreading rate of the moving front is determined by the propagation speed of this nonlinear traveling wave. Moreover, the capillary wave is unstable to the transverse disturbances, and subsequent evolution may give rise into a fascinating phenomenon of a fingering instability \citep{Huppert1982}, which depends on the free-surface profile at the onset of the destabilization. The capillary wave can be considered as a base state for the subsequent fingering. For this significance, it is important to study the quantitative capillary wave profile for understanding many features of film spreading and wetting.

For gravity driven coating flow, if the liquid film has high viscosity, the flow becomes low Reynolds number, and the inertia effect can be negligible where gravity, viscous force and surface tension dominate. A lubrication theory is widely used to model viscous flow and simplify the hydrodynamic equations \citep{Craster2009}. Numerous theoretical, experimental and numerical studies have examined the dynamics of gravity driven viscous films down an inclined plane \citep{Schwartz1989,Goodwin1991}. In these canonical problems, the profiles of capillary waves were well studied by analytical and high-resolution numerical methods in the framework of the lubrication theory. Subsequent researches were paying attention to film flow over more complex topography, such as flow down the outside of a horizontal \citep{Weidner1997,Evans2004} or vertical \citep{Smolka2011,Wray2013} cylinder. But most of these studies are based on numerical simulation, and the limited computational examples can not describe the commonality within these phenomena. Moreover, because of the multi-scale features in coating flow, developing an appropriate numerical method with small discretization errors is a challenging process. There are also some papers to give theoretical perspectives for coating flow on a general curved face \citep{Roy2002,Myers2002,Howell2003}. These studies put more emphasis on the mathematical formulation of governing equations in a general orthogonal curvilinear coordinate system. The lack of solution on specific surfaces makes these theories rather abstract, they would benefit from specific applications. One of the motivations in our study is to use the ideas of mathematical modeling behind these papers to solve more specific cases and find some similarities behind the phenomena.

If the fingering instability is not considered, one of the difficulties to theoretically describe the gravity driven capillary wave on curved surface is that the contact line may be a three-dimensional curve due to the non-symmetry of a general solid substrate, which is intractable compared to the straight contact line on inclined plane. Even so, there are typical symmetric geometries which are more common in nature. For example, cylindrical and spherical surfaces are the representatives with high symmetry and constant curvature. They can be considered as curved surfaces generated by translating or rotating a circle in three-dimensional space. These two similar curved surfaces can be considered as a class of problem and used as a starting point for more general study. The tangential component of the gravity varies as a sinusoidal function on both of the cylinder and sphere, due to the same base curve - a \emph{circle} for these two geometries, which is much different from the constant gravity component on inclined plane. Moreover, the curvature on cylindrical or spherical surface may affect the expression of the surface tension, which may make the free-surface profile essentially different from the planar problem. Some experiments concerning the flow and stability of thin films down the outside of a cylindrical and a spherical substrate had been examined by \citet{Takagi2010} recently. They also presented an analytical study of viscous current at the top of cylinder and sphere. But the theory is based on conditions that the film is constrained near the top where the streamwise gravity is approximated to increase linearly, and the surface tension is neglected everywhere for simplification. The main motivation in our study is to explain these phenomena from a more deductive perspective. We will derive the governing equations to model the leading order physics of the coating flow down the outside of a horizontal cylinder and a sphere. A regular perturbation method is used to expand the hydrodynamic equations and to obtain the leading order terms which is more tractable. If we assume that the initial film is located symmetrically at the top of the cylinder or sphere, the capillary wave formed later may become two-dimensional (for cylinder) or axisymmetric (for sphere) which makes the governing equation simplified further. It is well known that Rayleigh-Taylor instability may occur if the high-dense fluid is located above less-dense fluid. We do not consider moving contact line on the lower cylinder or hemisphere in this paper to avoid mixing two kinds of instabilities. The basic problems of coating flow may arise under two types of constraint conditions: constant volume or constant flux. This study focuses on the dynamics of coating flow which is formed by liquid of constant volume. We assume that the capillary wave is not to be destabilized by any disturbance, namely that, the fingering instability will never occur so that we can study long-term evolution of an unstable capillary wave, even though it may be difficult to realize in a laboratorial environment.

We will focus on the situations that a \emph{thin} film flows down the \emph{large} sized curved substrate. In this situation there is a small parameter in the evolution equation which makes the mathematics become a singular perturbation problem. Like the planar problem, the capillary wave on cylindrical or spherical surface always consists of three regions, each with its own characteristic scaling. In the \emph{outer} region where the free surface curvature is small, the body force that drives the liquid is resisted by the viscous force, and the surface tension plays a negligible role. In the \emph{inner} region local to the moving front, where the free surface curvature is large, the surface tension is balanced with body and viscous force, and a capillary ridge always evolves. A third \emph{contact} region exists at the advancing contact line, where the no-slip condition is no longer valid. In this paper, the general structures in both the outer and inner regions are studied. A method of matched asymptotic expansions can be used to match the outer profile and inner profile, which constitute a complete capillary wave. We will show that, the capillary waves on cylindrical and spherical surfaces are closely related to that on inclined plane, but still have their own characteristics.

The outline of this paper is as follows. In \ref{sec:theo}, the theoretical formulation is developed by two scalar Partial Differential Equations (PDE) that govern the coating flow of thin liquid films on cylindrical and spherical surfaces. Due to the symmetrically distributed initial conditions, the governing equation can be simplified to a two-dimensional form for cylindrical problem or an axisymmetric form for spherical problem. We use the precursor film condition with a compatible disjoining pressure model to simulate the partial wetting process near moving contact line. In \ref{sec:asym}, an asymptotic theory is elaborated to deal with the singular perturbation problem of high Bond number flow. In \ref{sec:compu}, the solution techniques for numerically solving the outer and inner equations introduced by the asymptotic theory and the complete evolution equations are discussed. In \ref{sec:numdis}, the numerical solutions of the outer equations, inner equations and the evolution equations are presented. We discuss the comparisons between the asymptotic theory and direct numerical solutions. The individuality and commonality in the two-dimensional and axisymmetric capillary waves are highlighted. Section \ref{sec:conrem} summarizes the conclusions of the present work.

\section{\label{sec:theo}Theoretical formulation}

\subsection{Governing equation on cylindrical surface}

\begin{figure}
  \centerline{\includegraphics{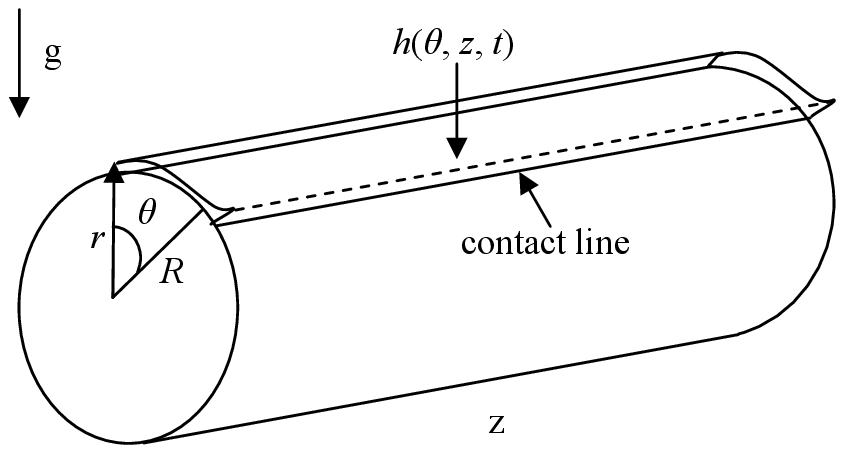}}
  \centerline{(a)}
  \centerline{\includegraphics{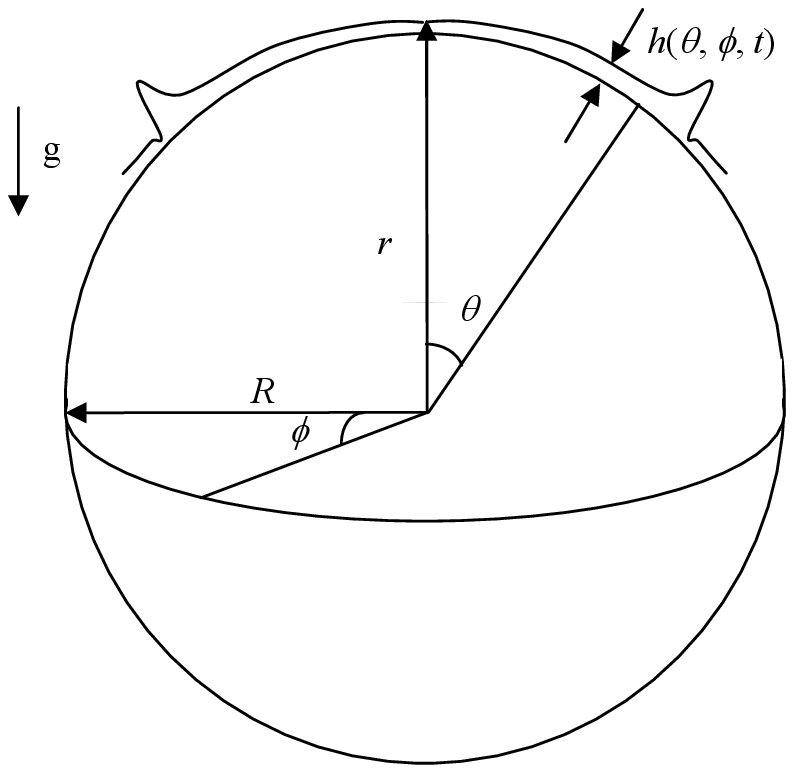}}
  \centerline{(b)}
  \caption{\label{fig:sketch} Sketch of thin liquid film spreading on upper (a) cylinder and (b) sphere.}
\end{figure}

An incompressible Newtonian liquid film of density $\rho$, viscosity $\mu$ and surface tension $\sigma$, flows under gravity on a stationary rigid cylinder of radius $R$, the axis of the cylinder is horizontal. The standard cylindrical coordinates of azimuthal angle $\theta$ and axial length $z$ are established on the cylindrical surface. The normal distance $n=r-R$ is defined as the third coordinate of the complete coordinate system, where $r$ is radial distance. The diagram of standard cylindrical coordinates and liquid film are sketched in Fig.~\ref{fig:sketch}(a), and note that the azimuthal angle $\theta$ measured downward from the top of the cylinder. The following scales are used to nondimensionalize the hydrodynamic equations in cylindrical coordinate system
\begin{subequations}
\begin{equation}\label{eq:scaleu}
u=Uu^{*}=\frac{\rho gH^{2}}{\mu}u^{*},v=Uv^{*},w=\epsilon Uw^{*}
\end{equation}
\begin{equation}\label{eq:scalep}
n=Hn^{*},R=LR^{*},t=\frac{L}{U}t^{*}=\frac{\mu L}{\rho gH^{2}}t^{*},p=Pp^{*}=\frac{\mu UL}{H^{2}}p^{*}
\end{equation}
\begin{equation}\label{eq:scalea}
z=Lz^{*}
\end{equation}
\end{subequations}
where $(u, v, w)$ are azimuthal, axial and radial components of the fluid velocity, $g$ is gravity acceleration, $t$ is time and $p$ is pressure, the variables with asterisk represent dimensionless. $L$ and $H$ denote length and thickness scales of the liquid film, respectively, and $\epsilon=H/L$ is aspect ratio. The values of velocity scale $U$ and pressure scale $P$ are standard Stokes scales when gravity and pressure terms on the right side of hydrodynamic equations are supposed to be in equilibrium with the viscous term.

A regular perturbation method is used to simplify the three-dimensional hydrodynamic equations when the aspect ratio, $\epsilon$, and reduced Reynolds number, $\epsilon^2\uRe(\uRe=\rho UL/\mu)$, are both sufficiently small. This ``long wave'' assumption is similar to a lubrication theory for coating flow on inclined plane, and the main difference is that the kind of solid substrate is extended to a curved surface. By substituting (\ref{eq:scaleu}), (\ref{eq:scalep}) and (\ref{eq:scalea}) into hydrodynamic equations, the simplified Stokes equations and continuity equation on cylindrical surface can be derived after some algebraic manipulations
\begin{subequations}
\begin{equation}\label{eq:cyrsu}
\mu\frac{\partial^{2}u}{\partial n^{2}}=\frac{1}{R}\frac{\partial p}{\partial \theta}-\rho g\sin\theta+O(\epsilon,\epsilon^{2}\uRe)
\end{equation}
\begin{equation}\label{eq:cyrsv}
\mu\frac{\partial^{2}v}{\partial n^{2}}=\frac{\partial p}{\partial z}+O(\epsilon,\epsilon^{2}\uRe)
\end{equation}
\begin{equation}\label{eq:cyrsw}
\frac{\partial p}{\partial n}=O(\epsilon,\epsilon^{3}\uRe)
\end{equation}
\begin{equation}\label{eq:cyrsc}
\frac{1}{R}\frac{\partial u}{\partial \theta}+\frac{\partial v}{\partial z}+\frac{\partial w}{\partial n}=O(\epsilon)
\end{equation}
\end{subequations}
The terms of $O(\epsilon)$ and $O(\epsilon^2\uRe)$ are perturbation terms in the framework of a perturbation method, and the leading order form can be obtained by neglecting all higher order terms. In this paper, only the terms explicitly expressed in the equations (\ref{eq:cyrsu})-(\ref{eq:cyrsc}) are retained to study the leading order physics. These leading order terms are transformed back to the dimensional forms for deriving a dimensional governing equation.

The boundary conditions to solve (\ref{eq:cyrsu})-(\ref{eq:cyrsc}) are no slip and no permeation conditions on solid substrate
\begin{equation}\label{eq:bcu}
u=v=w=0\mid_{n=0}
\end{equation}
the continuous conditions of normal and shear stress at the free surface of liquid film
\begin{subequations}
\begin{equation}\label{eq:bcn}
p=p_{s}+O(\epsilon)=-\sigma\kappa+O(\epsilon)\approx-\sigma(-\frac{1}{R}+\frac{1}{R^{2}}h+\frac{1}{R^{2}}\frac{\partial^{2}h}{\partial\theta^{2}}+
\frac{\partial^{2}h}{\partial z^{2}})+O(\epsilon)\mid_{n=h}
\end{equation}
\begin{equation}\label{eq:bct}
\mu\frac{\partial u}{\partial n}+O(\epsilon)=0,\mu\frac{\partial v}{\partial n}+O(\epsilon)=0\mid_{n=h}
\end{equation}
\end{subequations}
and the kinematic condition which holds the movement of the free surface
\begin{equation}\label{eq:bck}
w=\frac{\partial h}{\partial t}+\frac{u}{R}\frac{\partial h}{\partial \theta}+v\frac{\partial h}{\partial z}+O(\epsilon)\mid_{n=h}
\end{equation}
where $h(\theta,z,t)$ is the local thickness of liquid film. The Young-Laplace pressure $p_s$ is expressed in (\ref{eq:bcn}), and the mean curvature of free surface $\kappa$ on the right side is approximated by the summation of the mean curvature of cylindrical surface and the first order terms including $h$ and its derivatives introduced by additional film thickness \citep{Howell2003}. Note that the length scale to measure the Young-Laplace pressure is different from length scale $L$ in (\ref{eq:scalep}). A so-called capillary length which is derived by assuming the surface tension to be in equilibrium with the dominant driven force is often used to estimate the order of Young-Laplace pressure. In next section, we will show that the leading order physics may become singular in a special region if the surface tension is neglected, and the Young-Laplace pressure serves to smooth out singularities in the leading-order solution. Thus they are considered having a leading-order effect in the regions where the capillary length is much less than the global length scale $L$ (the long wave assumption holds because the capillary length is still greater than the global thickness scale $H$).

The azimuthal velocity $u$ and axial velocity $v$ can be determined by integrating (\ref{eq:cyrsu}) and (\ref{eq:cyrsv}) across the film thickness subject to (\ref{eq:bcu}) and (\ref{eq:bct}), the difference of radial velocity $w$ between solid substrate and free surface is obtained by integrating the continuity equation (\ref{eq:cyrsc}) along the normal direction. The final equation governing $h(\theta,z,t)$ is derived by substituting $w$ in (\ref{eq:bcu}) and (\ref{eq:bck}) into the integration of (\ref{eq:cyrsc})
\begin{equation}\label{eq:cydge}
\frac{\partial h}{\partial t}+\frac{1}{R}\frac{\partial}{\partial \theta}[\frac{h^{3}}{3\mu}(-\frac{1}{R}\frac{\partial p_{s}}{\partial\theta}+
\rho g\sin\theta)]+\frac{\partial}{\partial z}[\frac{h^{3}}{3\mu}(-\frac{\partial p_{s}}{\partial z})]=0
\end{equation}
The governing equation (\ref{eq:cydge}) can be nondimensionalized using the following scales
\begin{subequations}
\begin{equation}\label{eq:scaleh}
h=Hh^{*},t=\frac{3\mu R}{\rho gH^{2}}t^{*},p=\rho gRp^{*}
\end{equation}
\begin{equation}\label{eq:scalez}
z=Rz^{*}
\end{equation}
\end{subequations}
Then, the dimensionless form is obtained as
\begin{subequations}
\begin{equation}\label{eq:cyndge}
\frac{\partial h^{*}}{\partial t^{*}}+\frac{\partial}{\partial \theta}[h^{*3}(-\frac{\partial p_{s}^{*}}{\partial\theta}+\sin\theta)]+
\frac{\partial}{\partial z^{*}}(-h^{*3}\frac{\partial p_{s}^{*}}{\partial z^{*}})=0
\end{equation}
\begin{equation}\label{eq:cyndp}
p_{s}^{*}=-\frac{1}{Bo}(h^{*}+\frac{\partial^{2}h^{*}}{\partial\theta^{2}}+\frac{\partial^{2}h^{*}}{\partial z^{*2}})
\end{equation}
\end{subequations}
where the Bond number $Bo$ is
\begin{equation}\label{eq:bo}
Bo=\frac{\rho gR^{3}}{\sigma H}
\end{equation}
This dimensionless number is determined by density and surface tension of the liquid, gravity acceleration, radius of the cylinder and thickness scale of the liquid film. Note that the constant term in (\ref{eq:bcn}) which represents the mean curvature of the cylindrical surface is neglected in the dimensionless equation (\ref{eq:cyndp}) because the derivatives of constant curvature on cylindrical surface are zero in (\ref{eq:cydge}) and (\ref{eq:cyndge}).

\subsection{Governing equation on spherical surface}

The theoretical formulation of coating flow on spherical surface is analogous to the problem on cylindrical surface, if the standard spherical coordinates are used instead of cylindrical coordinates. Consider the gravity driven flow of a Newtonian liquid film with the same properties on a spherical substrate of radius $R$. The polar angle $\theta$, azimuthal angle $\phi$ and normal distance $n=r-R$ are introduced near the spherical surface. The sketch of the spherical substrate and the liquid film are shown in Fig.~\ref{fig:sketch}(b). The same scales expressed in (\ref{eq:scaleu}) and (\ref{eq:scalep}) are used to nondimensionalize the hydrodynamic equations in spherical coordinate system, and the simplified Stokes equations system can be extracted from the leading order
\begin{subequations}
\begin{equation}\label{eq:sprsu}
\mu\frac{\partial^{2}u}{\partial n^{2}}=\frac{1}{R}\frac{\partial p}{\partial \theta}-\rho g\sin\theta+O(\epsilon,\epsilon^{2}\uRe)
\end{equation}
\begin{equation}\label{eq:sprsv}
\mu\frac{\partial^{2}v}{\partial n^{2}}=\frac{1}{R\sin\theta}\frac{\partial p}{\partial \phi}+O(\epsilon,\epsilon^{2}\uRe)
\end{equation}
\begin{equation}\label{eq:sprsw}
\frac{\partial p}{\partial n}=O(\epsilon,\epsilon^{3}\uRe)
\end{equation}
\begin{equation}\label{eq:sprsc}
\frac{1}{R\sin\theta}\frac{\partial}{\partial \theta}(\sin\theta u)+\frac{1}{R\sin\theta}\frac{\partial v}{\partial \phi}+
\frac{\partial w}{\partial n}=O(\epsilon)
\end{equation}
\end{subequations}
The boundary conditions for (\ref{eq:sprsu})-(\ref{eq:sprsc}) are identical to that in cylindrical problem, except the expression of the Young-Laplace pressure
\begin{equation}\label{eq:spylp}
p_{s}=-\sigma\kappa\approx-\sigma(-\frac{2}{R}+\frac{2}{R^{2}}h+
\frac{1}{R^{2}\tan\theta}\frac{\partial h}{\partial\theta}+\frac{1}{R^{2}}\frac{\partial^{2}h}{\partial\theta^{2}}+
\frac{1}{R^{2}\sin^{2}\theta}\frac{\partial^{2}h}{\partial\phi^{2}})
\end{equation}
and the kinematic condition
\begin{equation}\label{eq:spbck}
w=\frac{\partial h}{\partial t}+\frac{u}{R}\frac{\partial h}{\partial \theta}+\frac{v}{R\sin\theta}\frac{\partial h}{\partial \phi}+O(\epsilon)\mid_{n=h}
\end{equation}
at the free surface around the sphere.

Finally the governing equation for local film thickness on spherical surface is
\begin{equation}\label{eq:spdge}
\frac{\partial h}{\partial t}+\frac{1}{R\sin\theta}\frac{\partial}{\partial \theta}[\sin\theta\frac{h^{3}}{3\mu}(-\frac{1}{R}\frac{\partial p_{s}}{\partial
\theta}+\rho g\sin\theta)]+\\
\frac{1}{R\sin\theta}\frac{\partial}{\partial \phi}[\frac{h^{3}}{3\mu}(-\frac{1}{R\sin\theta}\frac{\partial p_{s}}{\partial \phi})]=0
\end{equation}
Using the scales described in (\ref{eq:scaleh}), the dimensionless form is
\begin{subequations}
\begin{equation}\label{eq:spndge}
\frac{\partial h^{*}}{\partial t^{*}}+\frac{1}{\sin\theta}\frac{\partial}{\partial \theta}[\sin\theta h^{*3}(-\frac{\partial p_{s}^{*}}{\partial \theta}+\sin\theta)]+\frac{1}{\sin\theta}\frac{\partial}{\partial \phi}(-\frac{h^{*3}}{\sin\theta}\frac{\partial p_{s}^{*}}{\partial \phi})=0
\end{equation}
\begin{equation}\label{eq:spndp}
p_{s}^{*}=-\frac{1}{Bo}(2h^{*}+\frac{1}{\tan\theta}\frac{\partial h^{*}}{\partial\theta}+
\frac{\partial^{2}h^{*}}{\partial \theta^{2}}+\frac{1}{\sin^{2}\theta}\frac{\partial^{2}h^{*}}{\partial\phi^{2}})
\end{equation}
\end{subequations}
The definition of Bond number $Bo$ is identical to that expressed in (\ref{eq:bo}). According to the expression of $Bo$, for the liquid films which have the same physical property, the high Bond number represents thinner film flow on larger sized solid surface, and conversely the low Bond number represents thicker film flow on smaller sized surface.

\subsection{Two-dimensional flow on cylinder and axisymmetric flow on sphere}

Now let us focus on a special case of coating flow on cylindrical surface. A condition of $\frac{\partial h}{\partial z}=0$ is assumed to describe uniform distribution along axial direction. This condition represents a two-dimensional flow on the cylinder and the governing equations (\ref{eq:cyndge}) and (\ref{eq:cyndp}) can degenerate into a two-dimensional form
\begin{equation}\label{eq:tdeq}
\frac{\partial h}{\partial t}+\frac{\partial}{\partial \theta}\{h^{3}[\frac{1}{Bo}\frac{\partial}{\partial \theta}(h+
\frac{\partial^{2}h}{\partial \theta^{2}})+\sin\theta]\}=0
\end{equation}
The asterisk signs are neglected hereinafter. The two-dimensional case can be realized by setting an appropriate initial condition for (\ref{eq:cyndge}). Consider an initial liquid film of flat dimensionless thickness $h_{\ui}=1$ (the initial film thickness is used as the thickness scale $H$), is located at the top of the cylinder where the absolute value of azimuthal angle $\theta$ is less than a preset value $\theta_{\ui}$. Because the range of azimuthal angle $\theta$ is from $0$ to $2\pi$, to achieve more simplification, assume that the distribution of initial film is symmetrical about the vertical plane, so that we can only study a semicircle region of $0\leq\theta\leq\pi$. A precursor film of dimensionless thickness $b$ is assumed ahead of the uniform film to overcome the contact line singularity. The whole surface is prewetted by the precursor film and the precursor thickness is presented as a part of initial condition, mathematically, it is
\begin{equation}\label{eq:cwic1}
\begin{aligned}
  h_{\ui}(\theta) & =1 \qquad 0\leq\theta\leq\theta_{\ui} \\
  h_{\ui}(\theta) & =b \qquad \theta_{\ui}<\theta\leq\pi
\end{aligned}
\end{equation}
The perturbation method may be invalid at $\theta_{\ui}$ because of the discontinuousness. A continuous initial profile can be used to approximate (\ref{eq:cwic1})
\begin{equation}\label{eq:cwic2}
h_{\ui}(\theta)=\frac{1+b}{2}-\frac{1-b}{2}\tanh[a(\theta-\theta_{\ui})] \qquad 0\leq\theta\leq\pi
\end{equation}
where $\tanh(x)$ represents a hyperbolic tangent function, $a$ is a coefficient for controlling the width of transition zone. The boundary conditions of (\ref{eq:tdeq}) are
\begin{equation}\label{eq:cwbc}
h_{\theta},h_{\theta\theta\theta}=0\mid_{\theta=0,\pi}
\end{equation}
in which the subscripts represent derivatives. It implies that the flux into the flow domain is zero and the fluid volume can keep constant. The length scale of initial film can be expressed as $L=R\theta_{\ui}$, thus $\theta_{\ui}$ has a lower limit due to the request of the perturbation method
\begin{equation}\label{eq:initfc1}
\epsilon=H/L\ll1\Rightarrow\theta_{\ui}\gg H/R
\end{equation}

For coating flow on spherical surface, a similar constraint condition $\frac{\partial h}{\partial\phi}=0$ can be used to simplify governing equation (\ref{eq:spndge}) and (\ref{eq:spndp}) to an axisymmetric form
\begin{equation}\label{eq:axeq}
\frac{\partial h}{\partial t}+\frac{1}{\sin\theta}\frac{\partial}{\partial \theta}\{\sin\theta h^{3}[\frac{1}{Bo}\frac{\partial}{\partial \theta}(2h+\frac{1}{\tan\theta}\frac{\partial h}{\partial\theta}+\frac{\partial^{2}h}{\partial \theta^{2}})+\sin\theta]\}=0
\end{equation}
It represents an axisymmetric flow on the sphere. The initial conditions and boundary conditions of (\ref{eq:axeq}) are identical to (\ref{eq:cwic1}), (\ref{eq:cwic2}) and (\ref{eq:cwbc}). The main difference is that the range of polar angle $\theta$ is natively from $0$ to $\pi$, and the additional symmetric plane is not necessary for spherical problem. The evolution equations (\ref{eq:tdeq}) and (\ref{eq:axeq}) are the focus of this paper.

\subsection{The wetting model}

The wetting process commonly exists in the coating flow problems. There are three cases in the region near the contact line: a prewetting case, a complete wetting case and a partial wetting case \citep{Gennes1985}. For prewetting cases, there is not a physically real contact line (the whole solid surface is prewetted by a macroscopic liquid layer), and the contact area can be defined using the region where the bulk liquid meets the prewetted layer. If the solid surface is sufficiently dry in the region where the coating liquid has not arrived, the contact angle at the contact line may be zero or finite (depend on the particular liquid-solid material system), corresponding to complete or partial wetting cases. The film thickness near contact line may be microscopic (at the nanoscale) in complete or partial wetting cases. To approximate the combined microscopic intermolecular forces, a disjoining pressure model introduced by Frumkin and Derjaguin \citep{Churaev1995} relates observed equilibrium contact angles to the intermolecular forces that become important for liquid films of submicroscopic dimensions in partial wetting cases. The energy density associated with the disjoining pressure $\Pi$ is minimized when the film thickness assumes a very small value. A computationally convenient model function is
\begin{equation}\label{eq:ddjp}
\Pi=B[(\frac{h_{\mathrm{min}}}{h})^{n}-(\frac{h_{\mathrm{min}}}{h})^{m}]
\end{equation}
where the exponents $n$ and $m$ are positive constants with $n>m>1$. The constant $B$ is positive and has the dimension of pressure. The first term in (\ref{eq:ddjp}) represents liquid-solid repulsion while the second term is attractive, leading to a stable film thickness at $h_{\mathrm{min}}$. This stable film is assumed on the whole solid surface, which can be considered as a precursor film and is compatible with the precursor condition. In the lubrication limit where the equilibrium contact angles $\theta_{\ue}$ is small, the force balance is used to evaluate the constant $B$ \citep{Schwartz1998}
\begin{equation}\label{eq:djpco}
B=\frac{(n-1)(m-1)}{2h_{\mathrm{min}}(n-m)}\sigma\theta_{\ue}^{2}
\end{equation}
As a result, $\Pi$ is characterized by $h_{\mathrm{min}}$, $\theta_{\ue}$, $m$, and $n$ only. $B$ has nonzero value only in partial wetting cases, because the equilibrium contact angles $\theta_{\ue}$ is zero in complete wetting cases. Unlike Young-Laplace pressure, the value of disjoining pressure at any location depends only on $h$. $\Pi$ is zero when $h=h_{\mathrm{min}}$ and becomes vanishingly small for $h\gg h_{\mathrm{min}}$. Gradients of $\Pi$ drive liquid motion but the effect is only important in the immediate vicinity of apparent contact lines.

Since the disjoining pressure is assumed to depend on the local interfacial separation only, with no slope and curvature contributions, the validity of expressions (\ref{eq:ddjp}) and (\ref{eq:djpco}) also requires the small free-surface slope and substrate curvature approximation. The formula of disjoining pressure is independent on the type of curved substrate. In the present formulation, disjoining pressure is an additional interfacial effect that may be thought of as a modification to the Young-Laplace pressure
\begin{equation}\label{eq:totip}
p_{t}=p_{s}-\Pi
\end{equation}
where $p_{t}$ is the total interfacial pressure. The governing equations on cylindrical and spherical surface can be modified after replacing $p_{s}$ by $p_{t}$ in boundary condition (\ref{eq:bcn}). The dimensionless form of disjoining pressure is
\begin{equation}\label{eq:nddjp}
\Pi^{*}=\frac{1}{\epsilon^{2}Bo}\frac{(n-1)(m-1)}{2b(n-m)}\theta_{\ue}^{2}[(\frac{b}{h^{*}})^{n}-(\frac{b}{h^{*}})^{m}]
\end{equation}
where $\epsilon=H/R$ is the aspect ratio if the radius $R$ is used as the length scale, $b=h_{\mathrm{min}}/H$ is dimensionless thickness of the stable film which can be considered as dimensionless precursor thickness described previously. The evolution equations become
\begin{equation}\label{eq:tddjpeq}
\frac{\partial h}{\partial t}+\frac{\partial}{\partial \theta}\{h^{3}[\frac{1}{Bo}\frac{\partial}{\partial \theta}(h+
\frac{\partial^{2}h}{\partial \theta^{2}}+\Pi_{\ud})+\sin\theta]\}=0
\end{equation}
for cylinder, and
\begin{equation}\label{eq:axdjpeq}
\frac{\partial h}{\partial t}+\frac{1}{\sin\theta}\frac{\partial}{\partial \theta}\{\sin\theta h^{3}[\frac{1}{Bo}\frac{\partial}{\partial \theta}(2h+\frac{1}{\tan\theta}\frac{\partial h}{\partial\theta}+\frac{\partial^{2}h}{\partial \theta^{2}}+\Pi_{\ud})+\sin\theta]\}=0
\end{equation}
for sphere, respectively, where $\Pi_{\ud}$ is
\begin{equation}\label{eq:djpd}
\Pi_{\ud}=\frac{1}{\epsilon^{2}}\frac{(n-1)(m-1)}{2b(n-m)}\theta_{\ue}^{2}[(\frac{b}{h})^{n}-(\frac{b}{h})^{m}]
\end{equation}
which can be derived from (\ref{eq:nddjp}).

\section{\label{sec:asym}Asymptotic theory for high bond number flow}

Because the Bond number $Bo$ is a unique dimensionless number in two-dimensional equation (\ref{eq:tdeq}) and axisymmetric equation (\ref{eq:axeq}), the free-surface profile may depend on $Bo$ in different ways. The flow characteristics are distinguished by the value of Bond number. Under high Bond number condition, the evolution may become a singular perturbation problem. We use a method of matched asymptotic expansions to study the high Bond number limit of (\ref{eq:tdeq}) and (\ref{eq:axeq}). This method is firstly discussed by \citet{Moriarty1991} for modeling unsteady spreading of a liquid film on a vertical plate with small surface tension.

\subsection{\label{subsec:cyout}Outer equation on cylindrical surface}

If the Bond number is very large, i.e.\ $Bo\gg1$, then the terms including $Bo^{-1}$ can be neglected in an outer region where $h_{\theta\theta\theta}\ll Bo$, and (\ref{eq:tdeq}) is simplified to an outer form
\begin{equation}\label{eq:cyout}
\frac{\partial h}{\partial t}+\frac{\partial}{\partial \theta}(\sin\theta h^{3})=0
\end{equation}
The quasi-linear form of (\ref{eq:cyout}) is
\begin{equation}\label{eq:cyqlout}
\frac{\partial h}{\partial t}+3\sin\theta h^{2}\frac{\partial h}{\partial \theta}=-\cos\theta h^{3}
\end{equation}
This first order PDE can be solved using the method of characteristics, but unfortunately the analytical solutions can not be expressed using elementary functions due to the existence of a first kind elliptic integral in solving process.

Even so, in a specific region where $\theta\ll1$, $\sin\theta\approx\theta\approx 0$ and $\cos\theta\approx1$, (\ref{eq:cyqlout}) can be simplified further
\begin{equation}\label{eq:cyrout}
\frac{\partial h}{\partial t}=-h^{3}
\end{equation}
This equation is definitely valid in $\theta\ll1$ region, moreover, in the region where $\theta<1$, (\ref{eq:cyrout}) can be considered as a leading order equation if the trigonometric coefficients in (\ref{eq:cyqlout}) are expanded using Taylor series. The dimensional form of outer equation (\ref{eq:cyout}) was firstly analyzed in \citet{Takagi2010} and they derived a long-term similarity solution in small $\theta$ region. Following their ideas, the general solution of dimensionless equation (\ref{eq:cyrout}) can be given directly using the characteristics method
\begin{equation}\label{eq:cyoutgsol}
h=[2t+\frac{1}{h_{\ui}^{2}(\theta)}]^{-\frac{1}{2}}
\end{equation}
$h_{\ui}(\theta)$ is the initial profile. We use uniform film with a precursor layer expressed in (\ref{eq:cwic1}) as initial condition, the solution is
\begin{subequations}
\begin{equation}\label{eq:cyoutsol1}
h=(2t+1)^{-\frac{1}{2}} \qquad \theta\leq\theta_{\uF}(t)
\end{equation}
\begin{equation}\label{eq:cyoutsol2}
h=(2t+\frac{1}{b^{2}})^{-\frac{1}{2}} \qquad \theta>\theta_{\uF}(t)
\end{equation}
\end{subequations}
$\theta_{\uF}$ is the location of moving front. Because the initial profile is independent on azimuthal angle $\theta$, according to (\ref{eq:cyoutsol1}), the film may evolve with uniform thickness. The profile of (\ref{eq:cyoutsol1}) and (\ref{eq:cyoutsol2}) is discontinuous at moving front, which can be considered as a shock wave due to the hyperbolic type of the outer equation.

The front location $\theta_{\uF}$ can be determined by the conservation of fluid volume. If $V$ is the dimensionless volume of that portion of the initial film lying above the precursor layer, i.e.,
\[V=\int_{0}^{\theta_{\ui}}(1-b)\ud\theta=(1-b)\theta_{\ui}\]
Then $\theta_{\uF}$ can be calculated by
\begin{subequations}
\begin{equation}\label{eq:cyvolf}
\int_{0}^{\theta_{\uF}}(h_{\uF}-b_{\uF})\ud\theta=V
\end{equation}
\begin{equation}\label{eq:cyhf}
h_{\uF}=(2t+1)^{-\frac{1}{2}} \qquad b_{\uF}=(2t+\frac{1}{b^{2}})^{-\frac{1}{2}}
\end{equation}
\end{subequations}
so that
\begin{equation}\label{eq:cythetaf}
\theta_{\uF}(t)=\frac{V}{h_{\uF}-b_{\uF}}=\frac{V}{(2t+1)^{-\frac{1}{2}}-(2t+\frac{1}{b^{2}})^{-\frac{1}{2}}}
\end{equation}
The front speed can be evaluated by differentiation of (\ref{eq:cythetaf})
\begin{equation}\label{eq:cydthetaf}
\dot{\theta}_{\uF}=\theta_{\uF}\frac{h_{\uF}^{3}-b_{\uF}^{3}}{h_{\uF}-b_{\uF}}=
\theta_{\uF}(h_{\uF}^{2}+h_{\uF}b_{\uF}+b_{\uF}^{2})
\end{equation}
This quantity represents the propagation speed of the shock wave. Note that the precursor layer thickness is decreasing with time according to (\ref{eq:cyoutsol2}), however, in most cases the flux in precursor layer thickness $[O(b_{\uF}^{3})]$ is much smaller than the flux in the bulk film $[O(h_{\uF}^{3})]$, the value of $b_{\uF}(t)$ in (\ref{eq:cythetaf}) and (\ref{eq:cydthetaf}) can be approximately the constant value of $b_{\uF}(0)$, viz. $b$. Even though the above solutions are derived under the condition $\theta\ll1$, the definitions of $\theta_{\uF}$, $h_{\uF}$ and $b_{\uF}$ are general on the whole upper cylinder if the solution of (\ref{eq:cyout}) is known. In Appendix~\ref{adx:exoutinn}, we will use the method of characteristics to derive an implicit form including the introduction of elliptic integral, which can be considered as the exact outer solutions on the whole upper cylinder.

\subsection{\label{subsec:spout}Outer equation on spherical surface}

The evolution equation in the outer region on spherical surface can be simplified from (\ref{eq:axeq}) by neglecting the $Bo^{-1}$ terms
\begin{equation}\label{eq:spout}
\frac{\partial h}{\partial t}+\frac{1}{\sin\theta}\frac{\partial}{\partial \theta}(\sin^{2}\theta h^{3})=0
\end{equation}
The quasi-linear form is
\begin{equation}\label{eq:spqlout}
\frac{\partial h}{\partial t}+3\sin\theta h^{2}\frac{\partial h}{\partial \theta}=-2\cos\theta h^{3}
\end{equation}
The only difference between outer equations on cylinder and sphere is twice the right term of (\ref{eq:cyqlout}) in (\ref{eq:spqlout}), which results in the appearance of a second kind elliptic integral of Weierstrass's form and makes the elementary expression of analytical solution impossible, as shown in Appendix~\ref{adx:exoutinn}.

Analogously, in the region of $\theta\ll1$, the second term on the left side of (\ref{eq:spqlout}) can be neglected, the simplest form is
\begin{equation}\label{eq:sprout}
\frac{\partial h}{\partial t}=-2h^{3}
\end{equation}
With the same initial profile, a shock wave solution can also be obtained
\begin{subequations}
\begin{equation}\label{eq:spoutsol1}
h=\frac{1}{2}(t+\frac{1}{4})^{-\frac{1}{2}} \qquad \theta\leq\theta_{\uF}(t)
\end{equation}
\begin{equation}\label{eq:spoutsol2}
h=\frac{1}{2}(t+\frac{1}{4b^{2}})^{-\frac{1}{2}} \qquad \theta>\theta_{\uF}(t)
\end{equation}
\end{subequations}
At a given time, the solution profile on both sides of front location $\theta_{\uF}$ is uniform, which is identical to the cylindrical problem.

The same volume conservation method can be used to determine $\theta_{\uF}$. Note that the formula to calculate the net volume $V$ on spherical surface (eliminating the coefficient $2\pi$) is different from that on cylindrical surface
\begin{equation}\label{eq:spvolf}
\int_{0}^{\theta_{\uF}}(h_{\uF}-b_{\uF})\sin\theta \ud\theta \approx \int_{0}^{\theta_{\uF}}(h_{\uF}-b_{\uF})\theta\ud\theta = V
\end{equation}
Then $\theta_{\uF}$ is
\begin{equation}\label{eq:spthetaf}
\theta_{\uF}(t)=(\frac{2V}{h_{\uF}-b_{\uF}})^{\frac{1}{2}}=
[\frac{4V}{(t+\frac{1}{4})^{-\frac{1}{2}}-(t+\frac{1}{4b^{2}})^{-\frac{1}{2}}}]^{\frac{1}{2}}
\end{equation}
Due to the addition of a term $\sin\theta$ in the net volume formula, the front location $\theta_{\uF}$ varies like $t^{1/4}$ at later time when the effect of initial condition can be neglected, which is different from $t^{1/2}$ in (\ref{eq:cythetaf}). But coincidently, the front speed is identical to that in cylindrical problem
\[\dot{\theta}_{\uF}=\theta_{\uF}(h_{\uF}^{2}+h_{\uF}b_{\uF}+b_{\uF}^{2})\]
The front speed $\dot{\theta}_{\uF}$ was obtained in the region $\theta\ll1$, no matter on cylindrical or spherical surface. Even though we can't give the explicit solutions for (\ref{eq:cyout}) and (\ref{eq:spout}) in the region of $\theta\sim 1$, there is a general method to compute the front speed on the whole upper surface using the values of $\theta_{\uF}$, $h_{\uF}$ and $b_{\uF}$. For cylindrical problem, according to (\ref{eq:cyout}), the net flux across the front is $\sin\theta_{\uF}(h_{\uF}^3-b_{\uF}^3)$, which must exactly balance the flux defined using the front speed, $\dot{\theta}_{\uF}(h_{\uF}-b_{\uF})$
\begin{equation}\label{eq:dthetaf}
\dot{\theta}_{\uF}=\sin\theta_{\uF}(h_{\uF}^{2}+h_{\uF}b_{\uF}+b_{\uF}^{2})
\end{equation}
For spherical problem, according to (\ref{eq:spout}), the net flux is $\sin^2\theta_{\uF}(h_{\uF}^3-b_{\uF}^3)$, and from another perspective, $\dot{\theta}_{\uF}\sin\theta_{\uF}(h_{\uF}-b_{\uF})$, so that
\[\dot{\theta}_{\uF}=\sin\theta_{\uF}(h_{\uF}^{2}+h_{\uF}b_{\uF}+b_{\uF}^{2})\]
which is equivalent to the cylindrical problem. In the mathematical viewpoint, (\ref{eq:dthetaf}) belongs to a Rankine-Hugoniot relation of a weak solution system, as pointed out by \citet{Howell2003}, which describes the relationship between the states on both sides of a shock wave. This Rankine-Hugoniot relation is mathematically consistent with the outer equations (\ref{eq:spout}) and (\ref{eq:spout}) and the volume conservation formulae (\ref{eq:cyvolf}) and (\ref{eq:spvolf}).

\subsection{Inner equations on cylindrical and spherical surfaces}

In previous subsection, we found some similarities of outer solutions on cylindrical surface and spherical surface, but the long-term evolution of these two kinds of shock waves is essentially different (with different scaling law). In this subsection, we will simultaneously discuss the evolution equations in the inner region for both cylindrical and spherical problems.

The outer solution is invalid near the moving front, because the profile of shock wave is singular at $\theta_{\uF}$. The steepening effects in the region near $\theta_{\uF}$ involve rapid and localized increases in the free surface curvature, and will be vigorously resisted by surface tension. In this inner region, high order derivatives included in surface tension terms are important. We can evaluate the width of the inner region by assuming the highest order derivative in (\ref{eq:tdeq}) and (\ref{eq:axeq}) to be $O(1)$ term, $h_{\theta\theta\theta}\sim Bo$, so that, the width $\Delta\theta\sim Bo^{-1/3}$. The parameter $Bo^{-1/3}$ can be used as the length scale of inner region. For high Bond number flow, the inner region is very small, and the outer region comprises most of the coating area of the fluid. For low Bond number flow, the inner region is large and it may be difficult to define an outer region, the surface tension effect is important in the whole flow domain. The former is focused on in present study.

For cylindrical problem, an inner coordinate system can be established using the transformation
\begin{equation}\label{eq:innco}
\xi=Bo^{\frac{1}{3}}[\theta-\theta_{\uF}(t)]
\end{equation}
The inner coordinate $\xi$ defines a stretched coordinate system moving with the front speed. Using (\ref{eq:innco}), the two-dimensional equation (\ref{eq:tdeq}) can be transformed into the following
\begin{equation}\label{eq:cyinnful}
Bo^{-1/3}\frac{\partial h}{\partial t}-\dot{\theta}_{\uF}\frac{\partial h}{\partial\xi}+
\frac{\partial}{\partial \xi}[h^{3}\sin(\theta_{\uF}+Bo^{-1/3}\xi)+h^{3}(Bo^{-2/3}\frac{\partial h}{\partial\xi}+
\frac{\partial^{3}h}{\partial\xi^{3}})]=0
\end{equation}
Compared to the condition $Bo\gg1$ in the outer region, a more strong condition $Bo^{1/3}\gg1$ is assumed in the inner region. To keep the leading order and neglect all $O(Bo^{-1/3})$ and $O(Bo^{-2/3})$ terms, (\ref{eq:cyinnful}) can be simplified as
\begin{equation}\label{eq:innerlo}
-\dot{\theta}_{\uF}\frac{\partial h}{\partial \xi}+
\frac{\partial}{\partial\xi}(h^{3}\sin\theta_{\uF}+h^{3}\frac{\partial^{3}h}{\partial\xi^{3}})=0
\end{equation}

For spherical problem, using the same transformation (\ref{eq:innco}), the axisymmetric equation (\ref{eq:axeq}) in the inner region becomes
\begin{multline}\label{eq:spinnful}
Bo^{-1/3}\frac{\partial h}{\partial t}-\dot{\theta}_{\uF}\frac{\partial h}{\partial \xi}+\frac{1}{\sin(\theta_{\uF}+Bo^{-1/3}\xi)}\frac{\partial}{\partial \xi}\{h^{3}\sin^{2}(\theta_{\uF}+Bo^{-1/3}\xi)+\\
h^{3}\sin(\theta_{\uF}+Bo^{-1/3}\xi)[2Bo^{-2/3}\frac{\partial h}{\partial\xi}+\\
Bo^{-1/3}\frac{\partial}{\partial \xi}(\frac{1}{\tan(\theta_{\uF}+Bo^{-1/3}\xi)}\frac{\partial h}{\partial\xi})+\frac{\partial^{3}h}{\partial\xi^{3}}]\}=0
\end{multline}
By neglecting all $O(Bo^{-1/3})$ and $O(Bo^{-2/3})$ terms, a leading order equation is
\[-\dot{\theta}_{\uF}\frac{\partial h}{\partial \xi}+\frac{\partial}{\partial \xi}(h^{3}\sin\theta_{\uF}+h^{3}\frac{\partial^{3}h}{\partial\xi^{3}})=0\]
which is identical to the cylindrical problem (\ref{eq:innerlo}) (the front speeds are equivalent as discussed in last subsection). Note that the time derivative in (\ref{eq:cyinnful}) or (\ref{eq:spinnful}) is $O(Bo^{-1/3})$ term, only space derivatives are contained in the leading order equations, thus quasi-steady solutions can be expected for both cylindrical and spherical problems. It implies that when $Bo^{1/3}\gg 1$, an observer in the inner system moving with the front will not perceive any time evolution of the capillary wave profile. The profile extends infinitely far downstream and upstream, and appears essentially flat far away from the front.

Because the leading order inner equations on cylindrical and spherical surface are equivalent, now (\ref{eq:innerlo}) can be used as a starting point to study the common characteristics of the cylindrical and spherical problems. We can integrate (\ref{eq:innerlo}) with respect to $\xi$
\begin{equation}\label{eq:inncom}
-\dot{\theta}_{\uF}h+h^{3}\sin\theta_{\uF}+h^{3}\frac{\partial^{3}h}{\partial\xi^{3}}=d
\end{equation}
The integration constant $d$ can be given by matching the profile onto $h\rightarrow h_{\uF}$ as $\xi\rightarrow -\infty$, so that, $d=h_{\uF}^{3}\sin\theta_{\uF}-\dot{\theta}_{\uF}h_{\uF}$. Furthermore, if the film thickness $h$ and inner coordinate $\xi$ are transformed using
\begin{subequations}
\begin{equation}\label{eq:innerh}
h=h_{\uF}h'
\end{equation}
\begin{equation}\label{eq:innerxi}
\xi=\frac{h_{\uF}^{1/3}}{\sin^{1/3}\theta_{\uF}}\xi'
\end{equation}
\end{subequations}
then by substituting $d$, (\ref{eq:dthetaf}), (\ref{eq:innerh}) and (\ref{eq:innerxi}) into (\ref{eq:inncom}), the equation become
\begin{equation}\label{eq:innode}
\frac{\partial^{3}h'}{\partial\xi'^{3}}=\frac{1+\delta+\delta^{2}}{h'^{2}}-\frac{\delta+\delta^{2}}{h'^{3}}-1
\end{equation}
where $\delta(t)=b/h_{\uF}(t)$ is determined by the outer solution can be considered as a relative precursor thickness, with the boundary conditions
\begin{subequations}
\begin{equation}\label{eq:innerbc1}
h'\rightarrow 1 \quad \xi'\rightarrow -\infty
\end{equation}
\begin{equation}\label{eq:innerbc2}
h'\rightarrow \delta \quad \xi'\rightarrow +\infty
\end{equation}
\end{subequations}
It can be written as a third order Ordinary Differential Equation (ODE) because time plays a parametric role as $\delta(t)$ in (\ref{eq:innode}). This boundary value problem is classic, firstly analyzed by \citet{Tuck1990} elaborately, and subsequently cited by many previous literatures \citep{Spaid1996,Bertozzi1997}. These references focused on the general description of the inner structure for the draining or coating flow problems on planar surface, but according to (\ref{eq:cyinnful}) and (\ref{eq:spinnful}), in the inner region the profiles on cylindrical and spherical surface can degenerate into the planar front when $Bo^{1/3}\gg1$.

It is not a surprising result because the terms introduced by the curved substrate in (\ref{eq:cyinnful}) and (\ref{eq:spinnful}) are all of $O(Bo^{-1/3})$ or $O(Bo^{-2/3})$. The inner region is sufficiently small compared to the outer region under large $Bo$ condition. An observer in the inner region can not perceive the arc geometry on cylindrical or spherical surface because sufficiently short arc can be considered as a line segment. If we regard a cross-sectional circle on cylindrical surface (or a longitude line on spherical surface) as a series of line segments with increasing inclined angles, then each line segment can represent an inner region. In this inner region, the streamwise gravity component is a constant and the capillary ridge is analogous to the inner profile on inclined plane with the same inclined angle. On the other hand, the effect of the curved substrate can be introduced by deriving first order and second order inner equations in the framework of the matched asymptotic expansions. They are higher order corrections under high Bond number condition and the forms are dependent on the types of the curved surface.

The complete relationship between the final inner coordinate $\xi'$ and original outer coordinate $\theta$ can be obtained from (\ref{eq:innco}) and (\ref{eq:innerxi})
\begin{equation}\label{eq:inncoful}
\xi'=\frac{\sin^{1/3}\theta_{\uF}}{h_{\uF}^{1/3}}Bo^{1/3}(\theta-\theta_{\uF})
\end{equation}
Alternatively, if the width of inner region is expressed as $R\Delta\theta$, (\ref{eq:inncoful}) can be derived using the well known capillary length \citep{Spaid1996} on inclined plane to nondimensionalize $R\Delta\theta$
\begin{equation}\label{eq:capleth}
l=h_{\mathrm{N}}(3Ca)^{-1/3}
\end{equation}
where $h_{\mathrm{N}}=h_{\uF}H$ is the dimensional front thickness in the outer region, $Ca=\mu U/\sigma=\rho g\sin\alpha h_{\mathrm{N}}^{2}/3\sigma$ is capillary number, $\alpha$ is inclined angle. Thus, the following formula is obtained as
\[\Delta\xi'=R\Delta\theta/l=\frac{\sin^{1/3}\theta_{\uF}}{h_{\uF}^{1/3}}(\frac{\rho gR^{3}}{\sigma H})^{1/3}\Delta\theta=
\frac{\sin^{1/3}\theta_{\uF}}{h_{\uF}^{1/3}}Bo^{1/3}\Delta\theta\]
This derivation gives a direct connection between cylindrical (or spherical) and planar problem. The high Bond number condition implies the capillary length is much smaller than the radius of cylinder or sphere, i.e., $l\ll R$. Since the width of capillary ridge $\Delta\xi'$ can be obtained for a given $\delta(t)$ in (\ref{eq:innode}), we can estimate the width of inner region via
\begin{equation}\label{eq:asymw}
\Delta\theta=\frac{h_{\uF}^{1/3}\Delta\xi'}{\sin^{1/3}\theta_{\uF}Bo^{1/3}}
\end{equation}
According to (\ref{eq:asymw}), the width of capillary ridge is proportional to the cubic root of front thickness $h_{\uF}$, and inversely proportional to the cubic root of $Bo$ and sine of front location $\theta_{\uF}$. Thus we obtain a 1/3-power law for the width of capillary ridge which is an analytic expression derived from asymptotic theory. The asymptotic theory can be validated using the width ratio between the inner region and the outer region
\begin{equation}\label{eq:asymc}
\Delta \theta/\theta_{\uF}=\frac{h_{\uF}^{1/3}\Delta\xi'}{\theta_{\uF}\sin^{1/3}\theta_{\uF}Bo^{1/3}} \ll 1
\end{equation}
This ratio decreases with time because of increasing $\theta_{\uF}$ and decreasing $h_{\uF}$. For a given $Bo$, if the front location $\theta_{\uF}$ is small (mostly arise at initial time), the ratio may approach or even be greater than 1, then the asymptotic theory may be invalid. But as time increases the asymptotic theory may become leading order valid. On the other hand, if the asymptotic theory is expected to be valid from the initial time, i.e., $\Delta\theta/\theta_{\ui}\ll 1 \Rightarrow \frac{\Delta\xi'}{\theta_{\ui}\sin^{1/3}\theta_{\ui}Bo^{1/3}}\ll 1$, under the condition $\theta_{\ui}\ll 1$ the initial polar angle $\theta_{\ui}$ should satisfy
\begin{equation}\label{eq:initfc2}
\theta_{\ui}\gg\Delta\xi'^{3/4}Bo^{-1/4}
\end{equation}
Compared to condition (\ref{eq:initfc1}), this is an additional constraint condition for initial film extent.

\subsection{\label{subsec:comps}Composite capillary wave}

A quasi-steady solution $h'[\xi',\delta(t)]$ in the inner region is obtained for each choice of the one-parameter $\delta(t)$. This inner solution can be rewritten in outer coordinates using the transformation (\ref{eq:inncoful}), so that
\[h_{\mathrm{inner}}=h_{\uF}h'[\frac{\sin^{1/3}\theta_{\uF}}{h_{\uF}^{1/3}}Bo^{1/3}(\theta-\theta_{\uF}),b/h_{\uF}]=h_{\uF}h'(\theta,t)\]
In a typical matched asymptotic expansions method, if we have obtained both the outer and inner solutions in leading order, the leading order composite solution over the entire flow domain can be constructed using a multiplicative composite expansion \citep{VanDyke1975}
\begin{equation}\label{eq:mulcomp}
h_{\mathrm{comp}}=h_{\mathrm{outer}}h_{\mathrm{inner}}/h_{\mathrm{inter}}
\end{equation}
$h_{\mathrm{inter}}$ is the intermediate solution that is common to both of the outer and inner regions. Upstream of the moving front, where $\theta\leq\theta_{\uF}$, $h_{\mathrm{inter}}=h_{\uF}$, and $h_{\mathrm{outer}}$ is the solution of (\ref{eq:cyout}) or (\ref{eq:spout}). Downstream of the front, where $\theta>\theta_{\uF}$, the common value between the inner and outer solutions is the precursor layer thickness, so that $h_{\mathrm{inter}}$ is equal to $b$. The composite solution is then
\begin{equation}\label{eq:compsol}
\begin{aligned}
  h_{\mathrm{comp}} & = h_{\mathrm{outer}}(\theta)h'(\theta,t) \qquad & \theta\leq\theta_{\uF}\\
  h_{\mathrm{comp}} & = h_{\uF}h'(\theta,t) \qquad & \theta>\theta_{\uF}
\end{aligned}
\end{equation}

The continuity of composite solution can be guaranteed using this constructed method. Note that (\ref{eq:innode}) is translational invariance in the $\xi'$ direction, because $\xi'$ is not included explicitly on the right side of (\ref{eq:innode}) and the boundaries are located infinitely far. \mbox{$h'[\xi'+\Delta\xi',\delta(t)]$} is also a solution for this boundary value problem. The free translational parameter $\Delta\xi'$ can be determined by volume conservation if the profile of composite solution is given
\[\int_{0}^{\pi}(h_{\mathrm{comp}}-b)\ud\theta=V\]
for cylinder and
\[\int_{0}^{\pi}(h_{\mathrm{comp}}-b)\sin\theta\ud\theta=V\]
for sphere. This constructed method can be used on both cylindrical and spherical surfaces to obtain a composite capillary wave.

\subsection{Addition of disjoining pressure}

The asymptotic theory discussed previously should be modified intuitively if the disjoining pressure term is added in evolution equations. Starting from (\ref{eq:tddjpeq}) and (\ref{eq:axdjpeq}), because in the outer region the film thickness $h\gg b$, like the surface tension terms, the disjoining pressure can be neglected in the outer region, and this term does not appear in the outer equations on cylindrical and spherical surface. Now pay attention to the inner coordinate system described in (\ref{eq:innco}), for example, transform (\ref{eq:tddjpeq}) in this inner coordinates
\begin{multline}\label{eq:cydjpinnful}
Bo^{-1/3}\frac{\partial h}{\partial t}-\dot{\theta}_{\uF}\frac{\partial h}{\partial\xi}+
\frac{\partial}{\partial \xi}[h^{3}\sin(\theta_{\uF}+Bo^{-1/3}\xi)+h^{3}(Bo^{-2/3}\frac{\partial h}{\partial\xi}+\\
\frac{\partial^{3}h}{\partial\xi^{3}}+Bo^{-2/3}\frac{\partial\Pi_{\ud}}{\partial\xi})]=0
\end{multline}
Note that
\[Bo^{-2/3}\Pi_{\ud}=\frac{1}{(Bo^{1/3}\epsilon)^{2}}\frac{(n-1)(m-1)}{2b(n-m)}\theta_{\ue}^{2}[(\frac{b}{h})^{n}-(\frac{b}{h})^{m}]\]
if we assume that $Bo^{1/3}\epsilon\sim1$, this term can be retained in the leading order inner equation
\begin{equation}\label{eq:inndjpcom}
-\dot{\theta}_{\uF}h+h^{3}\sin\theta_{\uF}+h^{3}\frac{\partial^{3}h}{\partial\xi^{3}}+
Bo^{-2/3}\frac{\partial\Pi_{\ud}}{\partial\xi}=d
\end{equation}
Using (\ref{eq:innerh}) and (\ref{eq:innerxi}), the final form of inner equation is
\begin{equation}\label{eq:inndjpode}
\frac{\partial^{3}h'}{\partial\xi'^{3}}=\frac{1+\delta+\delta^{2}}{h'^{2}}-\frac{\delta+\delta^{2}}{h'^{3}}-1-
\frac{K}{h'}[m(\frac{\delta}{h'})^{m}-n(\frac{\delta}{h'})^{n}]\frac{\partial h'}{\partial\xi'}
\end{equation}
where the dimensionless contact angle parameter is
\begin{equation}\label{eq:djpodeco}
K=\frac{1}{(Bo^{1/3}\epsilon)^{2}h_{\uF}^{4/3}\sin^{2/3}\theta_{\uF}}\frac{(n-1)(m-1)\theta_{\ue}^{2}}{2\delta(n-m)}=
(3Ca)^{-2/3}\frac{(n-1)(m-1)\theta_{\ue}^{2}}{2\delta(n-m)}
\end{equation}

Only the rightmost term which represents disjoining pressure is added compared to (\ref{eq:innode}). The same leading order inner equation on spherical surface can be obtained from the transformation of (\ref{eq:axdjpeq}). Equation (\ref{eq:inndjpode}) is identical to the two-dimensional steady-state ODE which is derived by \citet{Eres2000} for a gravity-driven draining film on a vertical plate, in which a similar disjoining pressure model is used. Even though in partial wetting cases, the inner equation on cylindrical and spherical surfaces can degenerate into a common form which includes the disjoining pressure as additional terms.

As discussed in \ref{sec:intro}, besides the outer and inner region, there is a third region existing at the advancing contact line, in which the characteristic scaling is much smaller than the inner region. The disjoining pressure is only operative in this \emph{contact} region. Strictly speaking, the asymptotic theory described above only makes the inner solution matching to the outer solution, and an additional matched asymptotic expansion \citep{Tuck1990} can be implemented between the inner region and the contact region. But in present study, the inner region and the contact region are merged into a complete \emph{inner} region and modeled using a unified inner equation.

\section{\label{sec:compu}Computational issues}

\subsection{Numerical techniques for outer and inner equations}

Even though there is a characteristics method to solve the outer equation (\ref{eq:cyout}) or (\ref{eq:spout}) analytically, the explicit expressions can not be written in elementary functions due to the existence of elliptic integrals as discussed in \ref{subsec:cyout} and \ref{subsec:spout}. Nevertheless, an implicit form can be derived to calculate exact solutions of (\ref{eq:cyout}) or (\ref{eq:spout}). The outer profile at a given time can be modeled using some nonlinear implicit expressions which includes the first kind (for cylindrical problem) or second kind (for spherical problem) incomplete elliptic integral. These expressions can be considered as a system of independent nonlinear algebraic equations and the Newton-Kantorovich's method can be used to obtain numerical solutions. The outer profile constructed using this numerical technique is exact because only local errors are introduced by the iterative method and the high precision numerical solutions can be guaranteed by controlling the residuals.

A classic shooting scheme \citep{Tuck1990} is used to solve (\ref{eq:innode}) (or (\ref{eq:inndjpode}) in partial wetting cases) and construct the inner solution. Numerical integration is performed using an adaptive fourth-order Runge-Kutta solver. Initial conditions are generated from an asymptotic equation valid far upstream where the uniform layer is only slightly perturbed, then the numerical profile in the inner system is constructed. The detailed numerical methods for outer and inner equations can be found in Appendix~\ref{adx:exoutinn}.

\subsection{Numerical techniques for two-dimensional and axisymmetric evolution equations}

It is useful to develop a numerical approach to solve the complete two-dimensional and axisymmetric evolution equations (\ref{eq:tdeq}) and (\ref{eq:axeq}) (or (\ref{eq:tddjpeq}) and (\ref{eq:axdjpeq}) in partial wetting cases), because the asymptotic theory may be invalid in the condition of relatively small Bond number. A time-marching finite difference method for typical parabolic PDE is used to solve (\ref{eq:tdeq}) and (\ref{eq:axeq}). The second order accuracy for both time and space can be achieved by applying a central difference scheme for spatial discretization and a Crank-Nicolson scheme for temporal discretization. The final expression forms a system of nonlinear algebraic equations, which are solved using an iterative Newton-Kantorovich's method.

When the Bond number is sufficiently large, the spatial scales in outer and inner region are very different. In addition, the space step must be somewhat smaller than the precursor thickness $b$, in dimensionless units, in order to maintain accuracy in the contact region where the front of capillary wave meets the precursor layer. In complete or partial wetting cases, the precursor thickness may be at the microscale or nanoscale, which requires the space step in the contact region to be of the same order. To capture the profile near apparent contact line and the effect of precursor thickness on the macroscopic capillary wave, the space step should be adapted according to the local slope and curvature of the free-surface profile. An effective \emph{r}-adaptive method is used with this finite difference scheme to reduce the total number of nodal points. We also have found that an adaptive time-stepping method can increase computational efficiency observably where the time step is adjusted dynamically using an explicit scheme based on a preset maximum permissible change of film thickness between each time step. Thus both the space step and the time step in this numerical marching method are adaptive. For more details, see Appendix~\ref{adx:numtdax}.

\section{\label{sec:numdis}Numerical solutions and discussion}

\subsection{\label{subsec:outsol}Outer and inner solutions on cylindrical and spherical surfaces}

In this subsection we focus on the prewetting and complete wetting cases in which the disjoining pressure term in (\ref{eq:inndjpode}) is negligible (degenerate into (\ref{eq:innode})). The solutions of outer equations (\ref{eq:cyout}) and (\ref{eq:spout}) are obtained via the method of characteristics. (\ref{eq:cwic1}) is used as the initial profile for the following computational examples, the initial front location $\theta_{\ui}$ is set to be equal to $\pi/16$. This initial condition will not be changed unless otherwise stated. Corresponding to the outer solution at a certain time, the quasi-steady solutions of (\ref{eq:innode}) in the inner region are integrated using the shooting method.

\begin{figure}
  \includegraphics[width=0.49\textwidth]{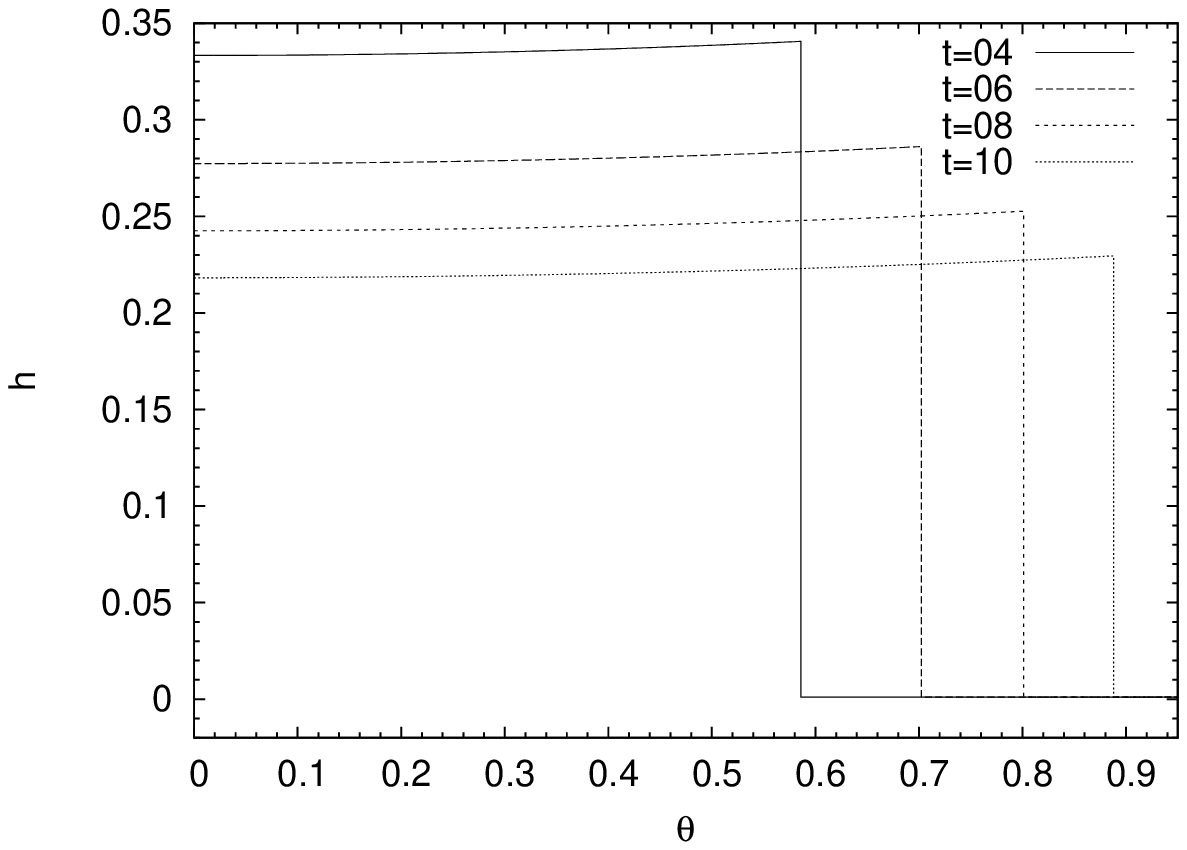}
  \includegraphics[width=0.49\textwidth]{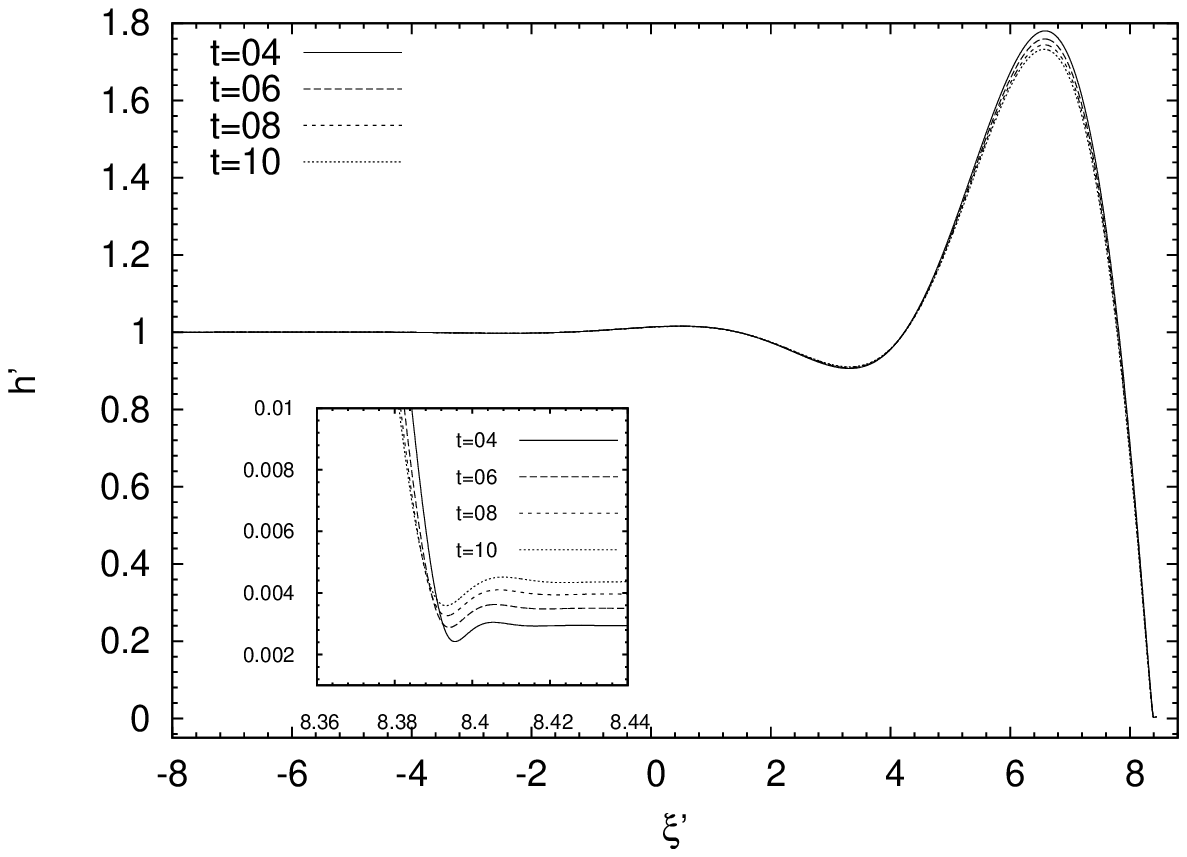}
  \centerline{(a)\hspace{0.45\textwidth}(b)}
  \caption{\label{fig:cyoutinnvst} Numerical profiles of (a) outer solutions and (b) inner solutions on cylindrical surface depending on time for the precursor thickness $b=0.001$. The inset in (b) is the refined profile near apparent contact line.}
\end{figure}

Figure~\ref{fig:cyoutinnvst}(a) shows the exact profiles of outer solutions on cylindrical surface depending on time for the precursor layer thickness $b=0.001$, the time is ranging from $t=4$ to $t=10$. A discontinuous shock wave is formed at all stages as displayed in this figure. In the region far away from the shock wave where $\theta$ is sufficiently small, the film thickness is approximately uniform, which is in accord with the asymptotic solution (\ref{eq:cyoutsol1}) given in \ref{subsec:cyout}. In the region where $\theta$ is near $O(1)$, it is worth our while to notice the monotonically increasing profiles which do not fit (\ref{eq:cyoutsol1}). Because asymptotic solution (\ref{eq:cyoutsol1}) is derived in the region of $\theta\ll1$, it may be invalid in the region where $\theta\sim1$. According to the coefficients in (\ref{eq:cyout}), the high degree terms in the Taylor series of the trigonometric functions may play an important role in $\theta\sim1$ region. As the substitute for an analytical solution in this region, the exact solution in Fig.~\ref{fig:cyoutinnvst}(a) can show the outlook of outer profile on the whole upper cylinder. If the front location $\theta_{\uF}\sim1$, it is believed that the corresponding front thickness $h_{\uF}$ is slightly greater than the value computed from (\ref{eq:cyoutsol1}). The numerical profiles of the inner solutions integrated from (\ref{eq:innode}) for different time are shown in Fig.~\ref{fig:cyoutinnvst}(b). The relative precursor thickness $\delta$ is calculated using $\delta=b/h_{\uF}$, where $b=0.001$, $h_{\uF}$ is obtained using the exact value in outer profiles at corresponding time as shown in Fig.~\ref{fig:cyoutinnvst}(a). A capillary ridge has been evolved near the moving front. The surface curvature is large at the capillary ridge, which represents the effect of the surface tension. Because $\delta$ is increasing with the decrease of front thickness as time increases, we can see the peak of the capillary ridge and the maximum slope at the moving front decreases with time. The inset in Fig.~\ref{fig:cyoutinnvst}(b) shows the refined profiles near apparent contact line, a primary minimum is typically observed in this contact region and the length scale is very small compared to the width of capillary ridge when $b=0.001$.

\begin{figure}
  \includegraphics[width=0.49\textwidth]{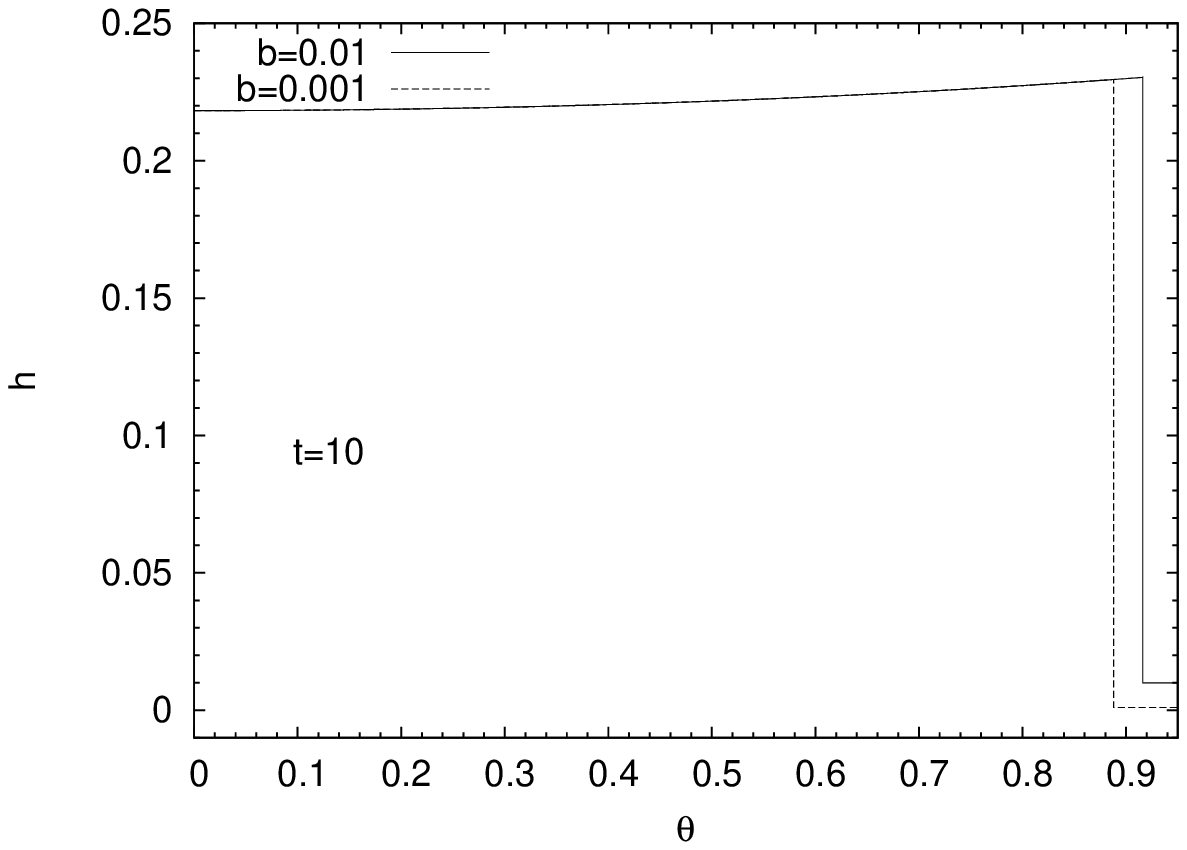}
  \includegraphics[width=0.49\textwidth]{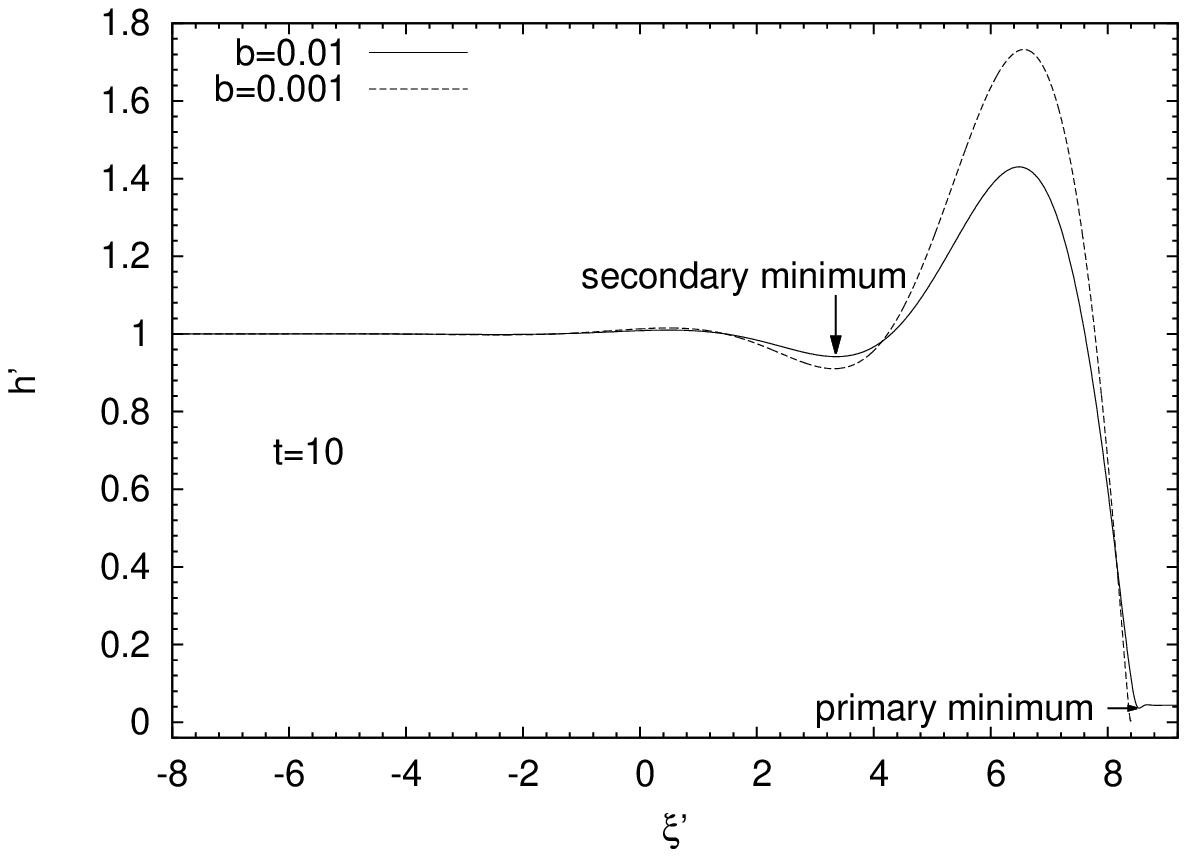}
  \centerline{(a)\hspace{0.45\textwidth}(b)}
  \caption{\label{fig:cyoutinnvsb} Comparison of (a) outer profiles and (b) inner profiles on cylindrical surface for different precursor thickness $b=0.01$ and $b=0.001$. The moment is at $t=10$.}
\end{figure}

As a reference compared with the complete wetting case, the precursor layer condition $b=0.01$ can be used to study a prewetting case. To present the influence of a relatively thick precursor layer on the bulk film thickness, Fig.~\ref{fig:cyoutinnvsb}(a) and \ref{fig:cyoutinnvsb}(b) show the direct comparison of the outer and inner solutions between $b=0.01$ profile and $b=0.001$ profile at $t=10$. The outer profile under condition $b=0.01$ is similar to that under condition $b=0.001$, except that the front location is greater than that in $b=0.001$ profile at the same time. The upstream $b=0.001$ profile coincide with the $b=0.01$ profile, and is cut at a smaller angular position by the shock wave. According to (\ref{eq:dthetaf}), the front speed is increasing with $b$, which can be seen clearly in Fig.~\ref{fig:cyoutinnvsb}(a). The peak of the capillary ridge and maximum slope at the moving front in the inner profile are the decreasing functions of $b$, which is presented obviously in Fig.~\ref{fig:cyoutinnvst}(b).

\begin{figure}
  \includegraphics[width=0.49\textwidth]{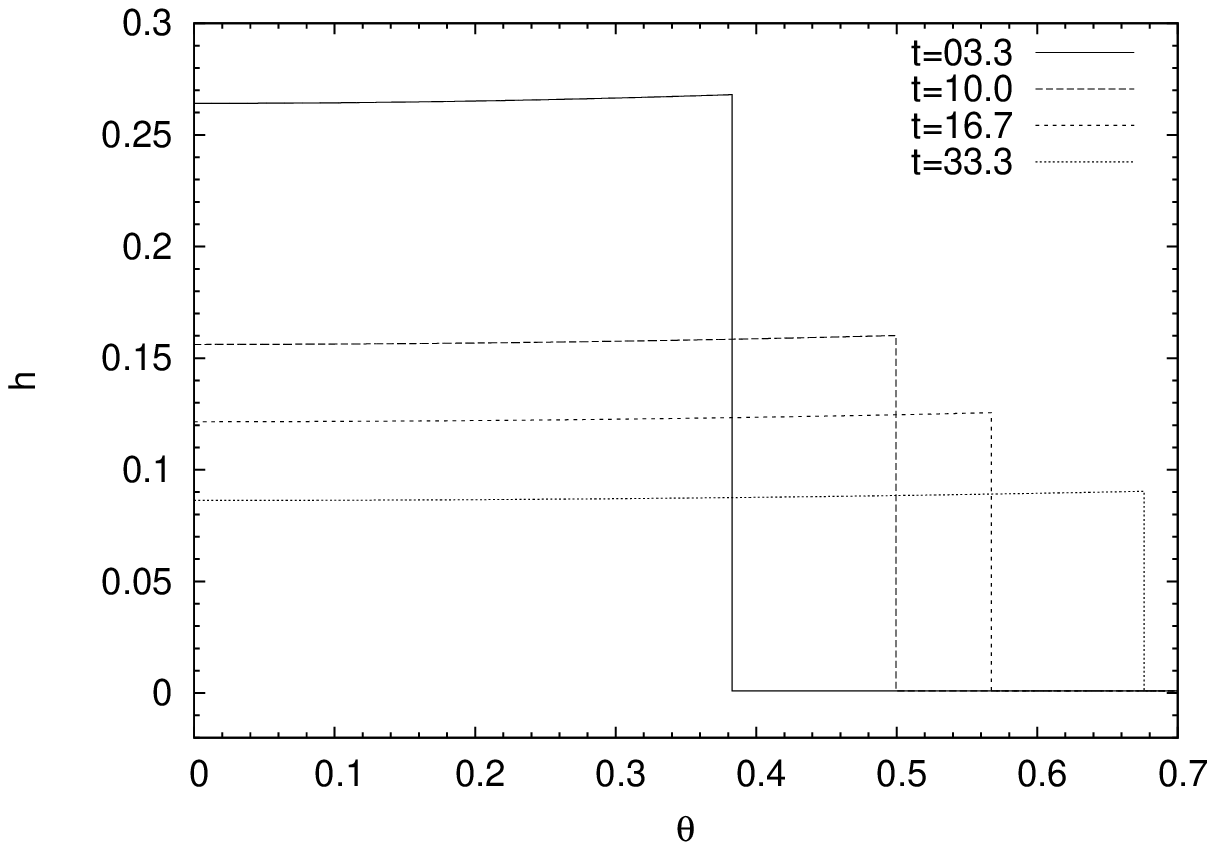}
  \includegraphics[width=0.49\textwidth]{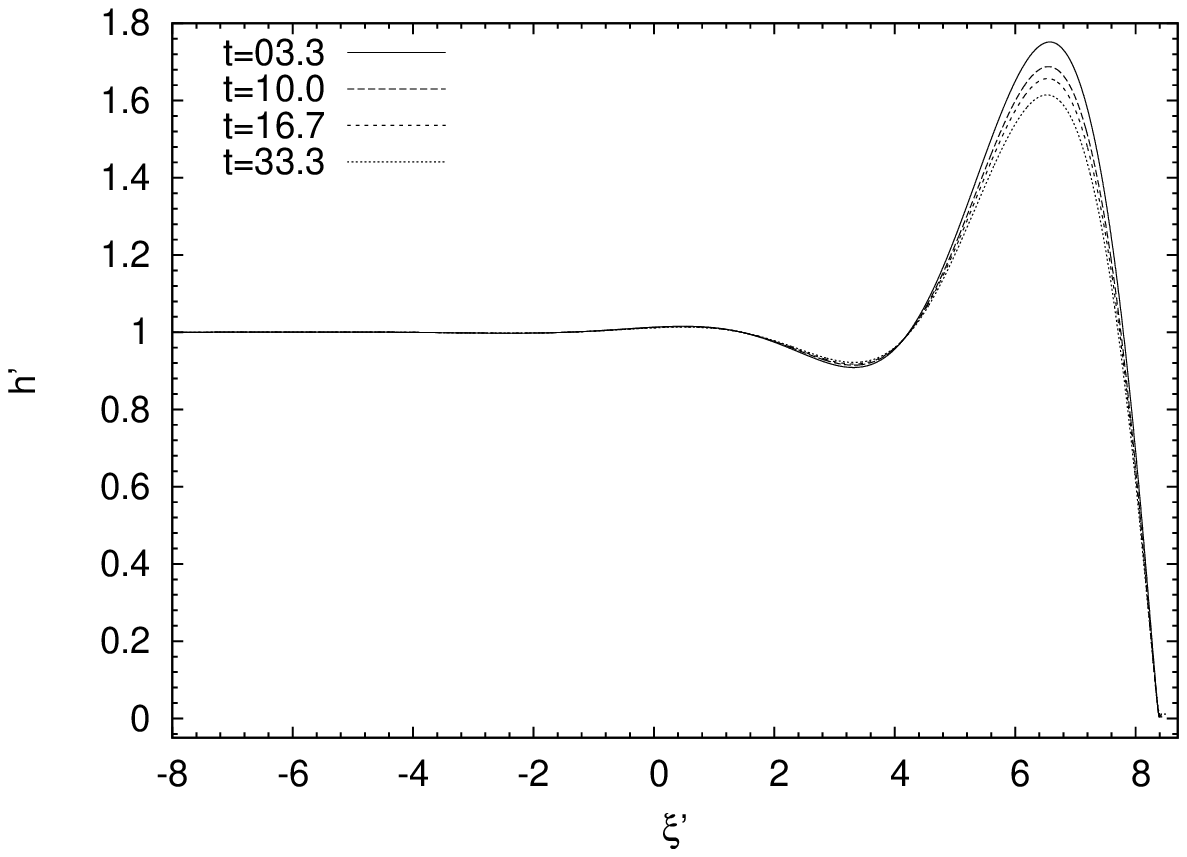}
  \centerline{(a)\hspace{0.45\textwidth}(b)}
  \caption{\label{fig:spoutinnvst} Numerical profiles of (a) outer solutions and (b) inner solutions on spherical surface depending on time for the precursor thickness $b=0.001$.}
\end{figure}

The numerical profiles of the outer solutions and corresponding inner solutions on spherical surface depending on time are shown in figures \ref{fig:spoutinnvst}(a) and \ref{fig:spoutinnvst}(b), respectively. The precursor thickness is $b=0.001$, and the time is ranging from $t=10/3$ to $t=100/3$. Even though the mathematical kernel of the outer equation (\ref{eq:spout}) is different from (\ref{eq:cyout}) according to Appendix~\ref{adx:exoutinn}, the typical feature of outer profiles on spherical surface is similar to the outer solutions on cylindrical surface (the profile in $\theta\ll1$ region is approximately uniform, the profile in $\theta\sim1$ region is monotonically increasing). But the front speed on spherical surface is observably slower than that on cylindrical surface (spent nearly sixfold time to arrive at the same angular position). Because the inner equation of spherical problem is identical to cylindrical problem (see (\ref{eq:innode})), compared to the inner profiles shown in Fig.~\ref{fig:cyoutinnvst}(b), the inner solutions in Fig.~\ref{fig:spoutinnvst}(b) are just another parametric family with different values of parameter $\delta$, which also belong to the one-parameter solution set of (\ref{eq:innode}).

\subsection{\label{subsec:innsol}The spreading of the moving front}

\begin{figure}
  \includegraphics[width=0.49\textwidth]{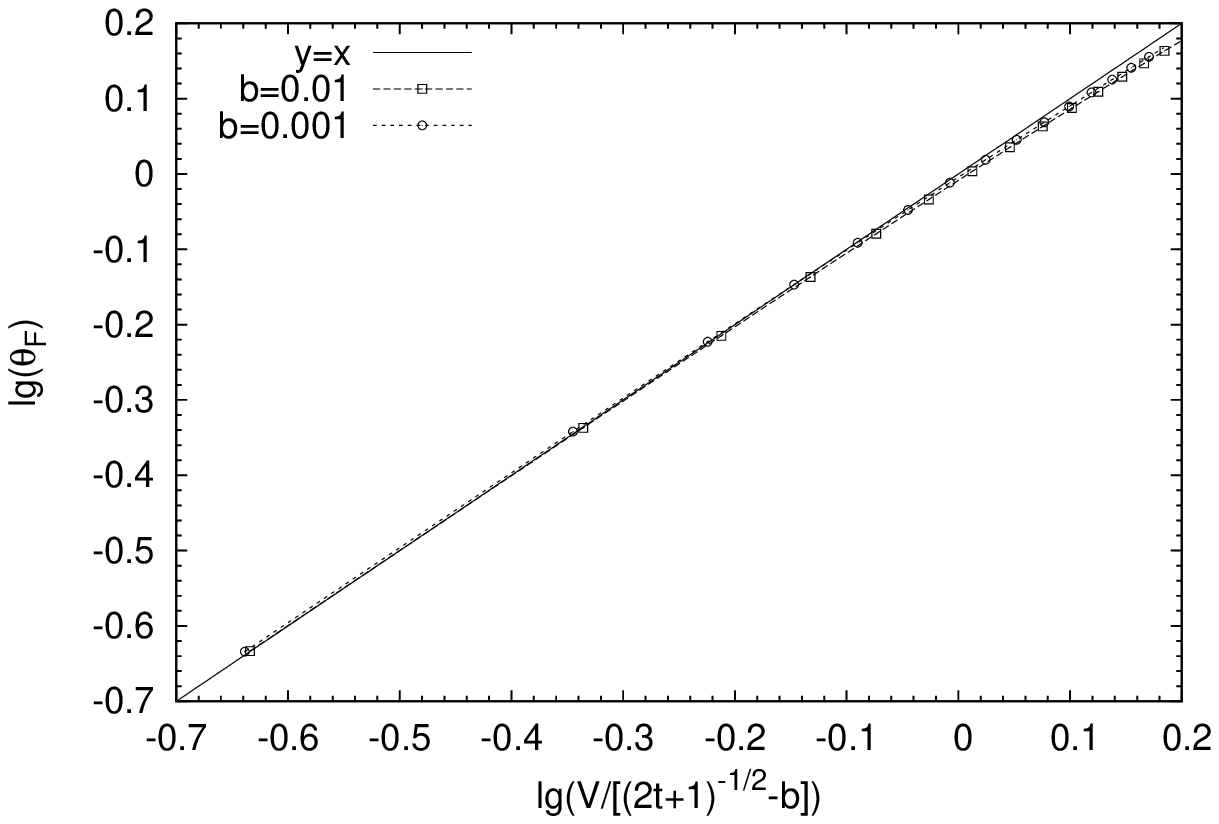}
  \includegraphics[width=0.49\textwidth]{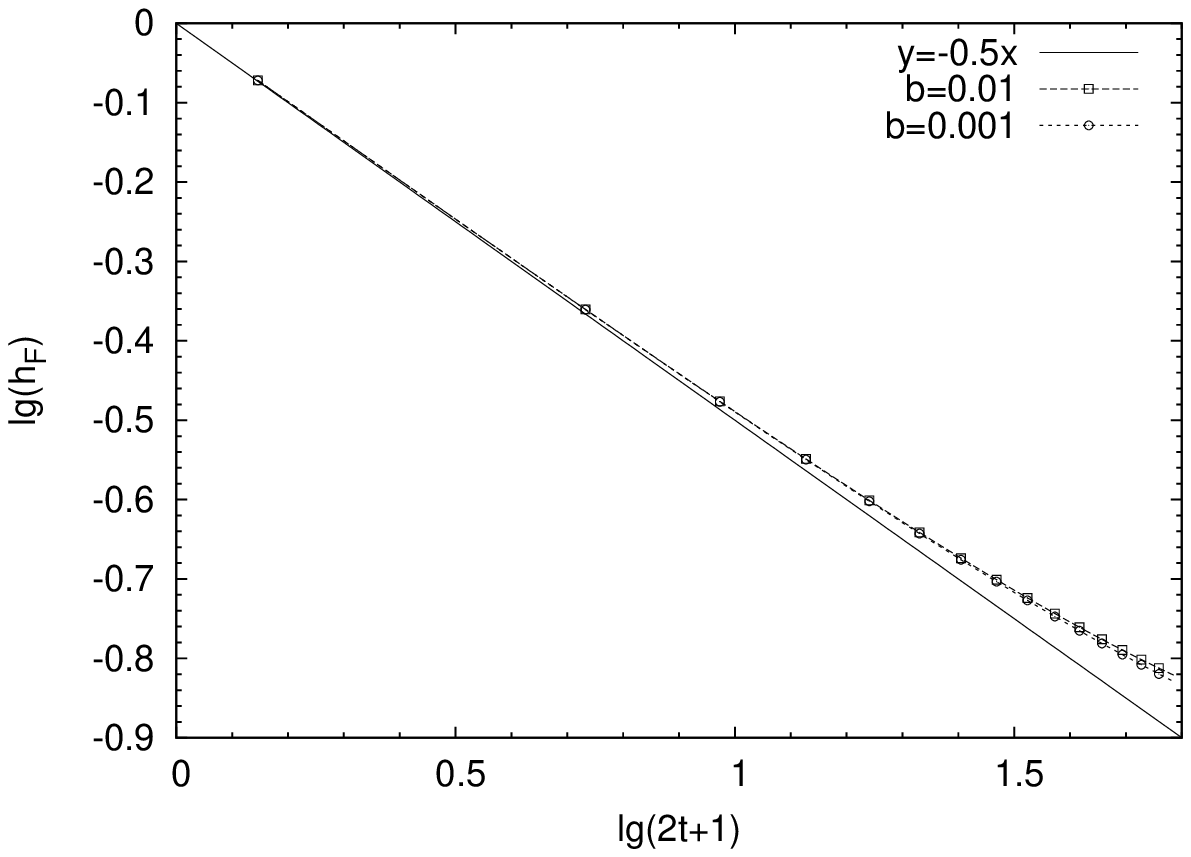}
  \centerline{(a)\hspace{0.45\textwidth}(b)}
  \caption{\label{fig:cythetafvst} (a) Log-log plots of front location versus time captured from the outer profiles on cylindrical surface. The precursor layer thickness is $b=0.01$ and $b=0.001$. The solid line $y=x$ is generated according to asymptotic formula (\ref{eq:cythetaf}) in top region. (b) Log-log plots of the front thickness versus time for precursor thickness $b=0.01$ and $b=0.001$ on cylindrical surface. The solid line $y=-0.5x$ is generated according to asymptotic solution (\ref{eq:cyoutsol1}).}
\end{figure}

In high Bond number flow, the spreading speed of the moving front (front speed) can be approximated by the propagation speed of the shock wave in the outer solution. The numerical relationship between the shock wave location and time is constructed by recording the front location which is calculated using the exact outer solution at a given time. Figure~\ref{fig:cythetafvst}(a) shows the log-log plot of the exact front location against time on the cylindrical surface for the precursor layer thickness $b=0.01$ and $b=0.001$ respectively. The time is ranging from $t=0.2$ to $t=30$ and transformed using the asymptotic formula (\ref{eq:cythetaf}) as the abscissa. The asymptotic relationship between the front location $\theta_{\uF}$ and time $t$ in $\theta\ll1$ region can be computed using (\ref{eq:cythetaf}). A straight line of $y=x$ which represents the analytical reference is also plotted. It is shown that the curves in small $\theta$ region agree well with the reference line, and the transient front location $\theta_{\uF}$ scales like $t^{1/2}$, which accords with the long-term scaling law in (\ref{eq:cythetaf}). In the $\theta\sim1$ region where (\ref{eq:cythetaf}) may be invalid, the slopes of numerical curves are slightly smaller than the reference line, which indicates a slower front speed due to the monotonically increasing film thickness. The monotonically increasing variation of the outer profile in $\theta\sim1$ region can be seen more clearly in Fig.~\ref{fig:cythetafvst}(b), which shows the log-log plots of the exact front thickness against time for precursor thickness $b=0.01$ and $b=0.001$. The solid line $y=-0.5x$ is generated according to asymptotic solution (\ref{eq:cyoutsol1}). As time increases, the front thickness in both of the $b=0.01$ case and $b=0.001$ case deviates from (\ref{eq:cyoutsol1}) obviously. The front thickness in $b=0.01$ case is slightly greater than that in $b=0.001$ case, because of the further front location where the $b=0.01$ profile arrives. In fact, as discussed in Appendix~\ref{adx:exoutinn} (see (\ref{eq:cyht})), the asymptotic solution (\ref{eq:cyoutsol1}) is just a particular solution of outer equation (\ref{eq:cyout}) at $\theta=0$. For $b=0.01$ case, according to the outer solution at $t=30$, when the front location $\theta_{\uF}$ arrives near $\pi/2$, the front thickness $h_{\uF}$ is about 18\% greater than the film thickness at $\theta=0$.

\begin{figure}
  \includegraphics[width=0.49\textwidth]{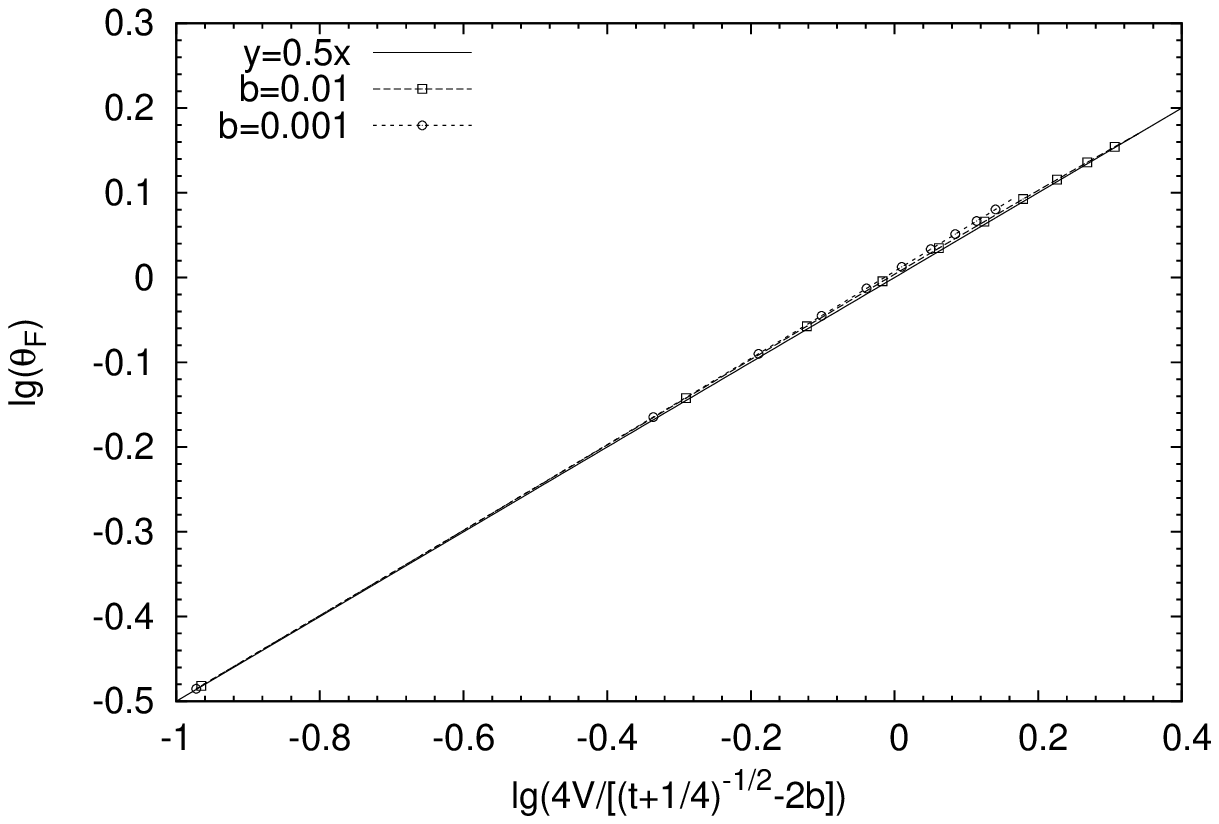}
  \includegraphics[width=0.49\textwidth]{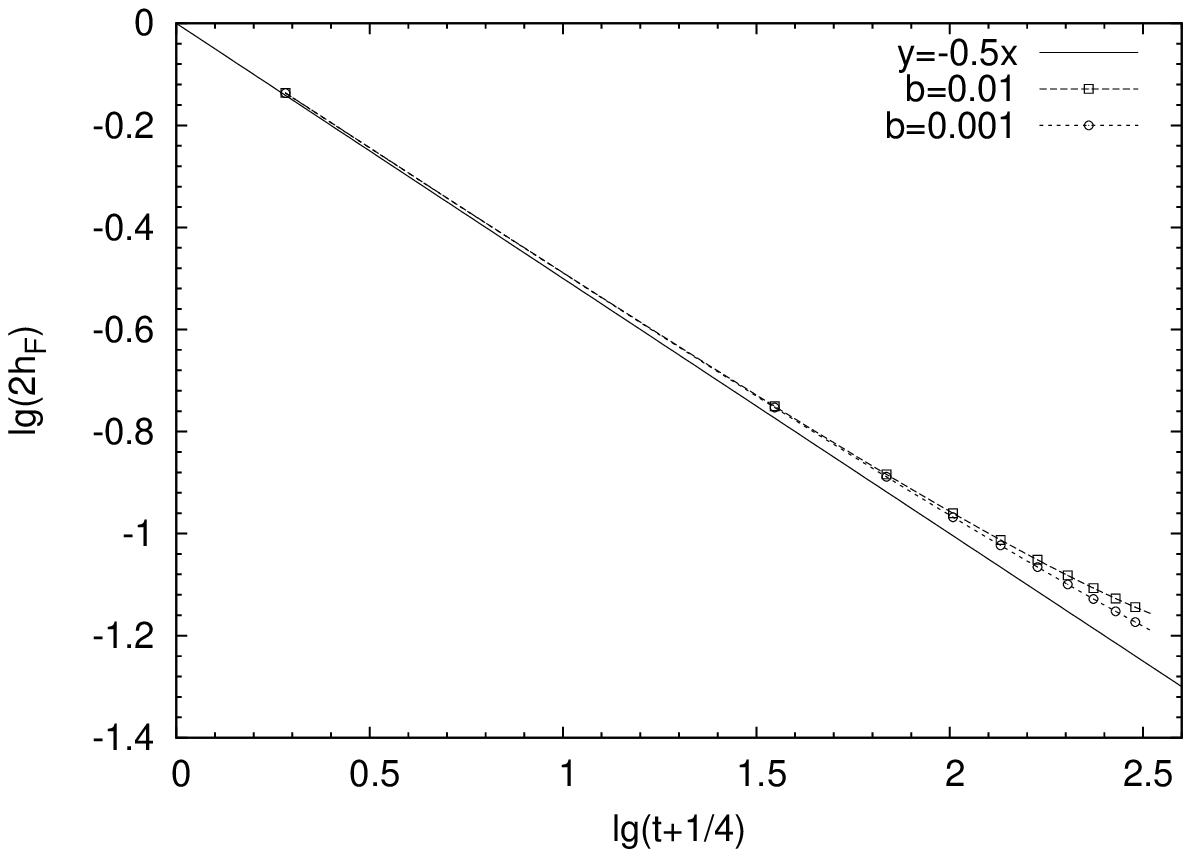}
  \centerline{(a)\hspace{0.45\textwidth}(b)}
  \caption{\label{fig:spthetafvst} (a) Log-log plots of front location versus time captured from the outer profiles on spherical surface. The precursor layer thickness is $b=0.01$ and $b=0.001$. The solid line $y=0.5x$ is generated according to asymptotic formula (\ref{eq:spthetaf}) in top region. (b) Log-log plots of front thickness versus time for precursor thickness $b=0.01$ and $b=0.001$ on spherical surface. The solid line $y=-0.5x$ is generated according to asymptotic solution (\ref{eq:spoutsol1})}
\end{figure}

Figure~\ref{fig:spthetafvst}(a) shows the log-log plots of the exact front location against time recorded from outer solutions on spherical surface. The precursor thickness is $b=0.01$ and $b=0.001$ and the time is ranging from $t=5/3$ to $t=1000/3$. The time is scaled using the asymptotic formula (\ref{eq:spthetaf}) as the abscissa. A straight line of $y=0.5x$ is also plotted as the reference line. It seems that the numerical curves agree well with the reference line in all stages (except that the curve of $b=0.001$ case is slightly greater than the reference line at later time), and the transient front location $\theta_{\uF}$ scales like $t^{1/4}$, which accords with the long-term scaling law in (\ref{eq:spthetaf}). But there are some coincidences in this comparison. Because the volume formula (\ref{eq:spvolf}) to calculate front location (\ref{eq:spthetaf}) assumes a linearly increasing $\theta$ on the whole upper hemisphere, it is not exact and the actual coefficient $\sin\theta$ in (\ref{eq:spvolf}) is sub-linear. The faster increasing $\theta$ results in a smaller front location which may cancel out the results of a greater front location induced by the effect of uniform profile given in (\ref{eq:spoutsol1}). In fact there are indeed more differences between the outer profile in $\theta\sim1$ region and the asymptotic solution in $\theta\ll1$ region for spherical problem. Figure~\ref{fig:spthetafvst}(b) shows the log-log plots of the exact front thickness against time for precursor thickness $b=0.01$ and $b=0.001$. The solid line $y=-0.5x$ is plotted according to asymptotic solution (\ref{eq:spoutsol1}). The similar deviations from the reference line for $b=0.01$ case and $b=0.001$ case are also observed in spherical problem. In extreme situation, according to the outer solution at $t=1000/3$ and $b=0.01$, when the front location $\theta_{\uF}$ arrives near $\pi/2$, the front thickness $h_{\uF}$ is about 27\% greater than the film thickness at $\theta=0$, which is more than that calculated in cylindrical problem (18\%).

\subsection{Comparison between composite solutions and direct solutions}

Because the Bond number $Bo$ does not appear in the outer equations (\ref{eq:cyout}) or (\ref{eq:spout}) and leading order inner equation (\ref{eq:innode}), the outer and inner profiles on both cylindrical and spherical surfaces are not affected by the value of $Bo$ directly. The action of $Bo$ is revealed after transforming the inner solution to the outer (original) space. In this subsection, we will use the matched method (\ref{eq:compsol}) discussed in \ref{subsec:comps} to construct the numerical profiles of composite solutions for different Bond number, which are the final descriptions of the complete capillary wave in the framework of the asymptotic theory. For comparison purposes, the direct numerical solutions of complete evolution equations (\ref{eq:tdeq}) and (\ref{eq:axeq}) (direct solutions for short) are also calculated by the adaptive numerical marching method given in Appendix~\ref{adx:numtdax}. These numerical profiles can be used as powerful benchmarks to examine the validation of the asymptotic theory. Since the Bond number of most computational examples implemented here is a large number, we define a logarithmic Bond number $logBo=\lg(Bo)$ for convenience of expression.

\begin{figure}
  \includegraphics[width=0.49\textwidth]{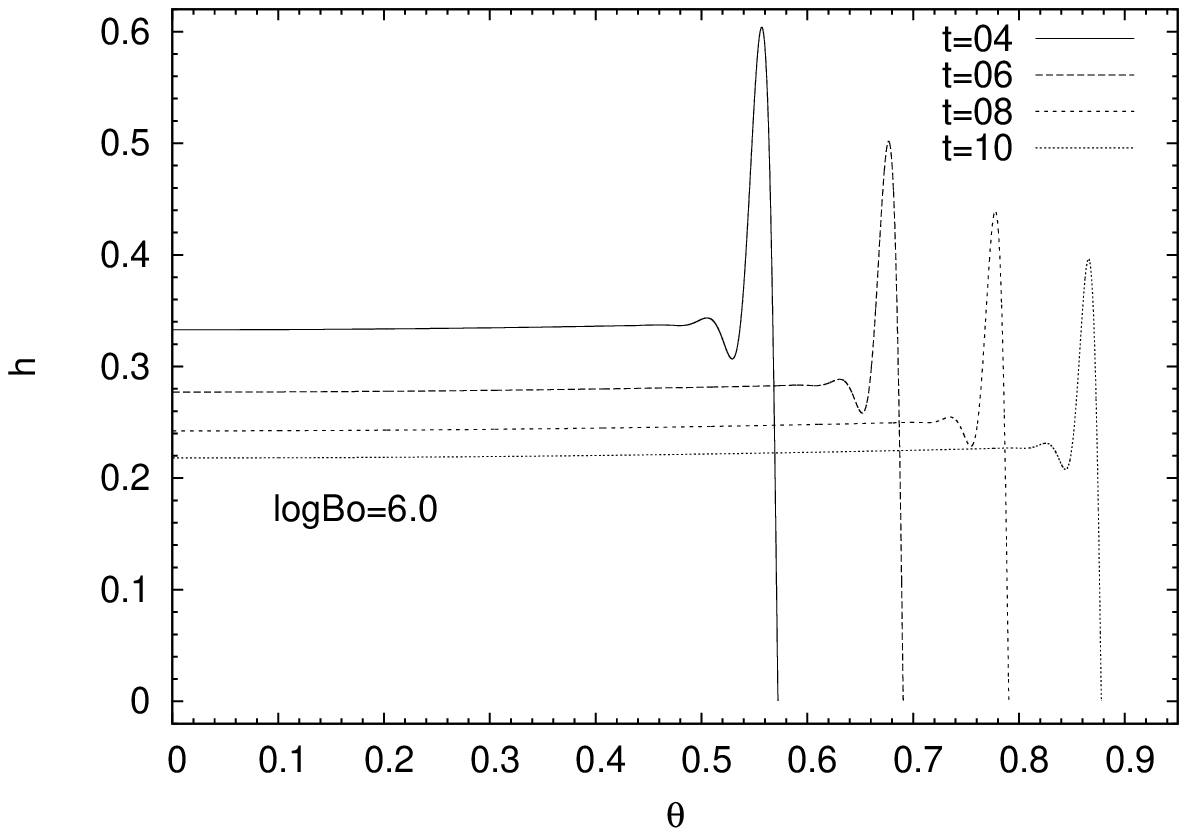}
  \includegraphics[width=0.49\textwidth]{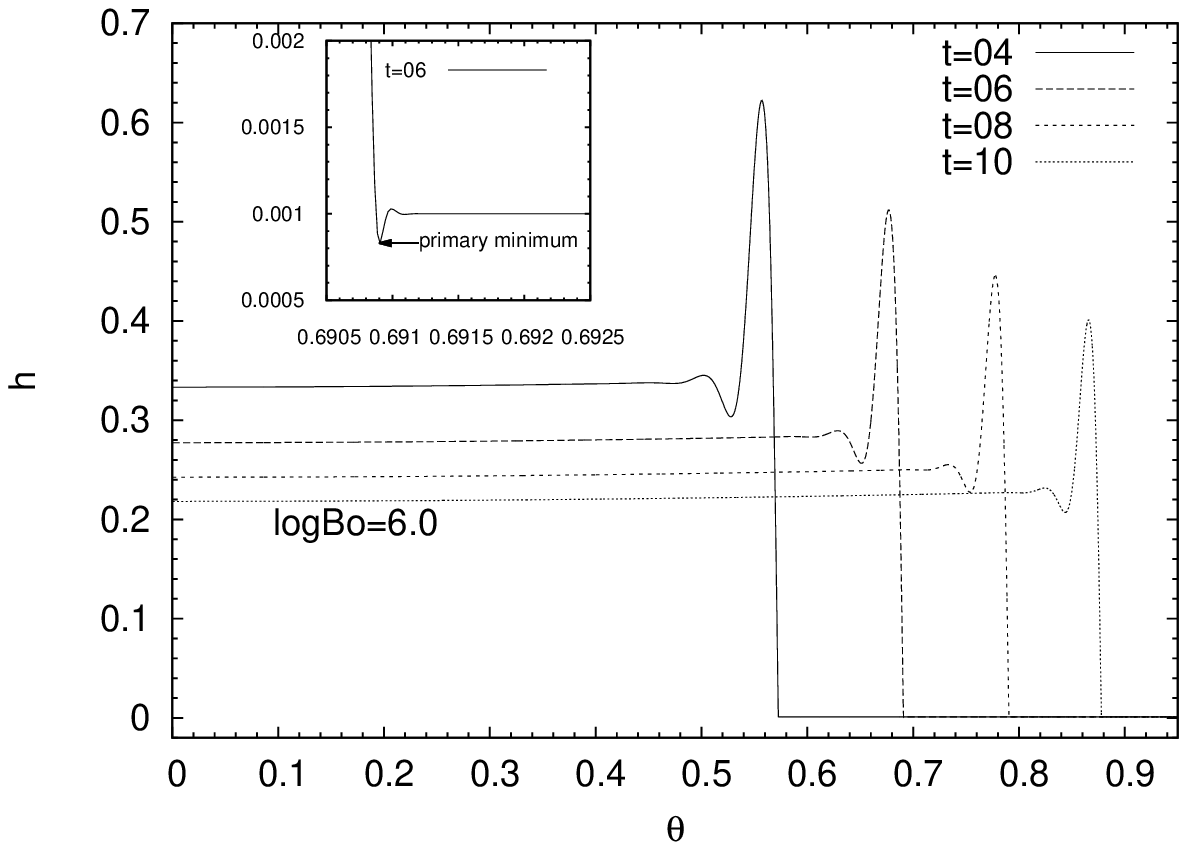}
  \centerline{(a)\hspace{0.45\textwidth}(b)}
  \caption{\label{fig:cycompvst} Numerical profiles of (a) composite solutions and (b) direct solutions on cylindrical surface depending on time. The logarithmic Bond number is $logBo=6.0$ and the precursor thickness is $b=0.001$. The inset in (b) is the refined profile near apparent contact line.}
\end{figure}

The composite solutions on cylindrical surface can be constructed by merging the outer profiles shown in Fig.~\ref{fig:cyoutinnvst}(a) and inner profiles shown in Fig.~\ref{fig:cyoutinnvst}(b) for a given Bond number. Figure~\ref{fig:cycompvst}(a) shows the numerical profiles of these composite solutions for a typical large Bond number $Bo=1\times10^6$, corresponding to a logarithmic Bond number $logBo=6.0$, and the precursor thickness is $b=0.001$. The outer and inner profiles are merged into a complete capillary wave in the whole flow domain. Under this Bond number, the width of the inner region (where the composite curve is dominated by the inner profile) is not large as shown in Fig.~\ref{fig:cycompvst}(a), and the basic assumption (\ref{eq:asymc}) for the asymptotic theory can be satisfied. Figure~\ref{fig:cycompvst}(b) shows the typical profiles of direct solutions depending on time under the conditions of $logBo=6.0$ and $b=0.001$. These profiles are obtained from the direct numerical solutions of two-dimensional evolution equation (\ref{eq:tdeq}), and the initial condition is set using (\ref{eq:cwic2}). The range of time are identical to that shown in Fig.~\ref{fig:cycompvst}(a). The capillary ridge has been calculated directly from the evolution equation. It is apparent that the outer and inner region in these complete profiles can be distinguished. The outer region is far away from the capillary ridge where the free-surface derivative is small. The inner region can be defined at the capillary ridge where the curvature is observable. The width ratio between the inner and outer region is relatively large at early time, and decreasing as the time increases, which had been proved in (\ref{eq:asymc}). The inset in Fig.~\ref{fig:cycompvst}(b) shows the refined free-surface profile near apparent contact line at $t=6$, which benefits from the adaptive method for directly solving the evolution equation. A primary minimum displayed is similar to that in inner profiles as shown in Fig.~\ref{fig:cyoutinnvst}(b).

\begin{figure}
  \includegraphics[width=0.49\textwidth]{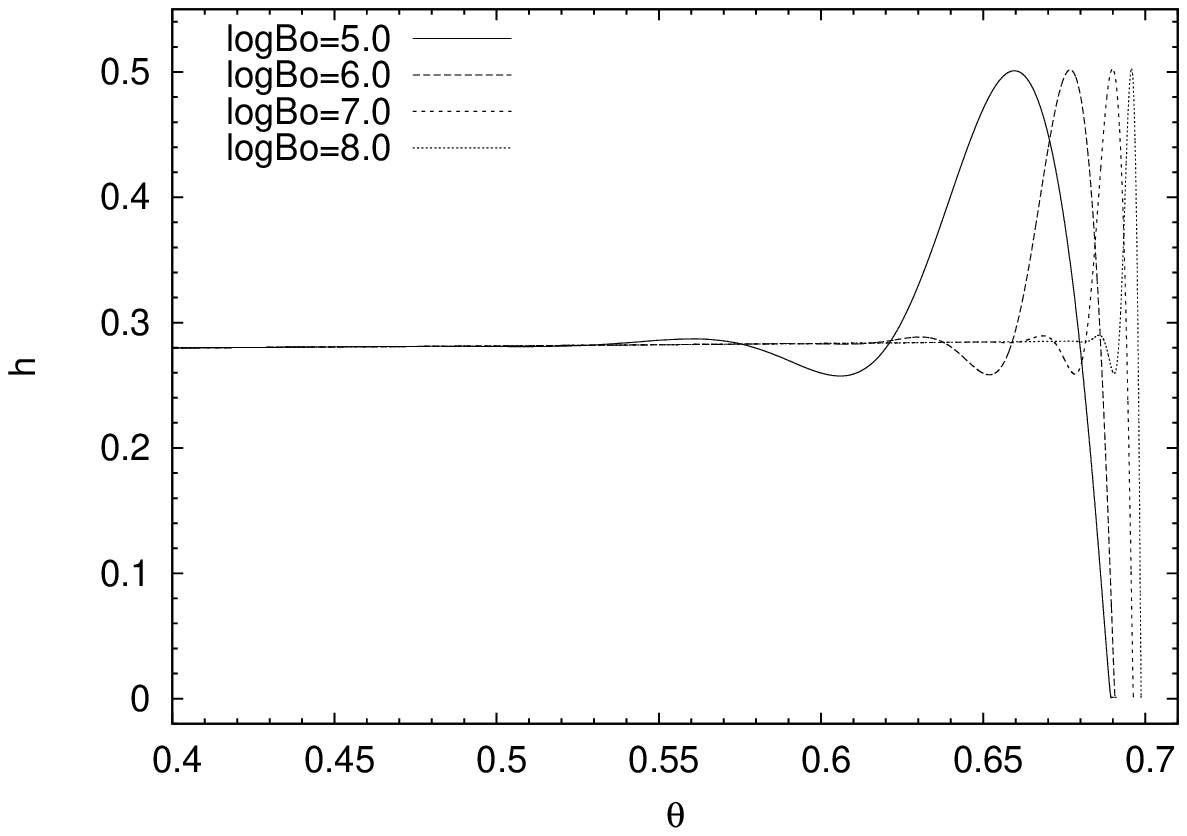}
  \includegraphics[width=0.49\textwidth]{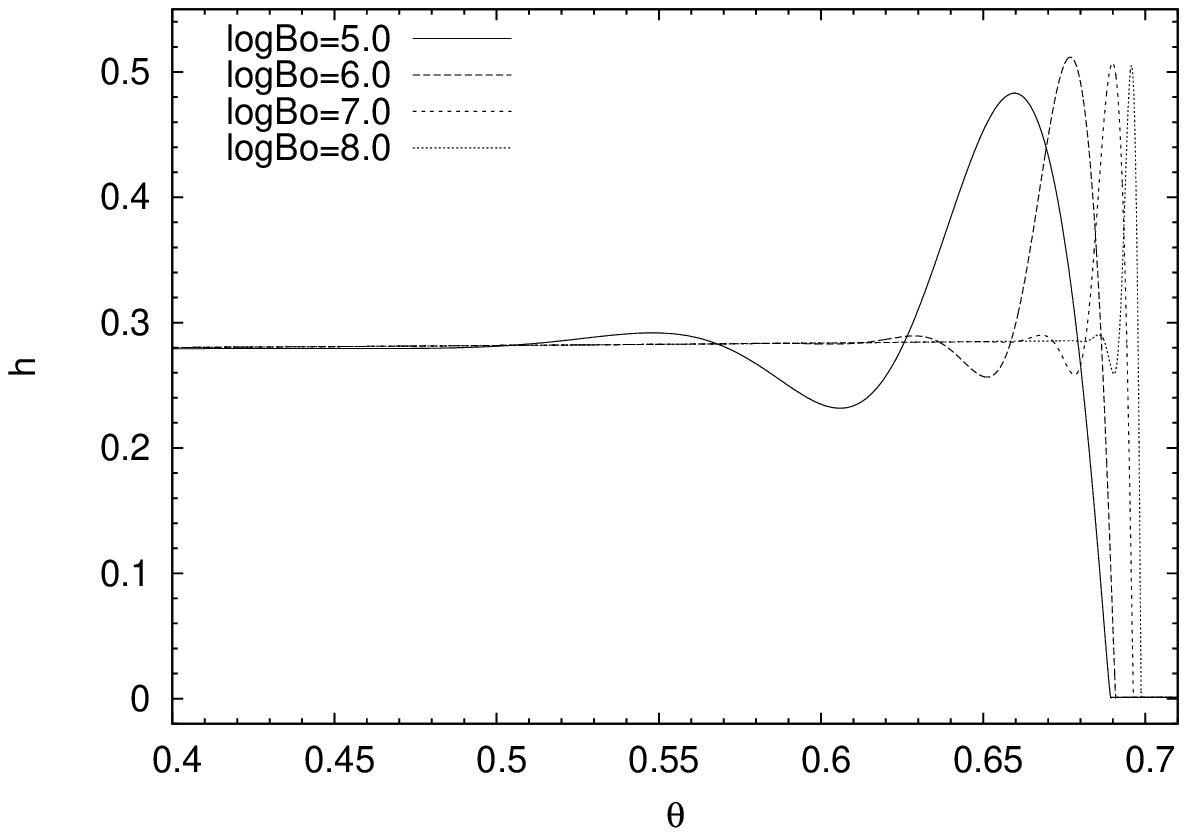}
  \centerline{(a)\hspace{0.45\textwidth}(b)}
  \caption{\label{fig:cycompvsbo} Numerical profiles of (a) composite solutions and (b) direct solutions on cylindrical surface depending on Bond number ranging from $10^5$ to $10^8$. The precursor thickness is $b=0.001$ and the time is at $t=6$.}
\end{figure}

To study the effect of the Bond number on the composite profiles, the composite solutions are constructed using different Bond number ranging from $logBo=5.0$ to $logBo=8.0$. Figure~\ref{fig:cycompvsbo}(a) shows the profiles of the composite solutions depending on $Bo$ at $t=6$. According to (\ref{eq:inncoful}), the smaller Bond number corresponds to wider inner region. It is clear in Fig.~\ref{fig:cycompvsbo}(a) that the width of the capillary ridge (inner region) decrease with $Bo$, but the maximum slope near the apparent contact line is an increasing function of $Bo$. As comparative references, the typical profiles of complete capillary wave in the same Bond number range at the same time as Fig.~\ref{fig:cycompvsbo}(a) are shown in Fig.~\ref{fig:cycompvsbo}(b). In these direct solutions, the width ratio between the inner and outer region is decreasing obviously as $Bo$ increases, which convincingly verifies the asymptotic condition \ref{eq:asymc}.

\begin{figure}
  \includegraphics[width=0.49\textwidth]{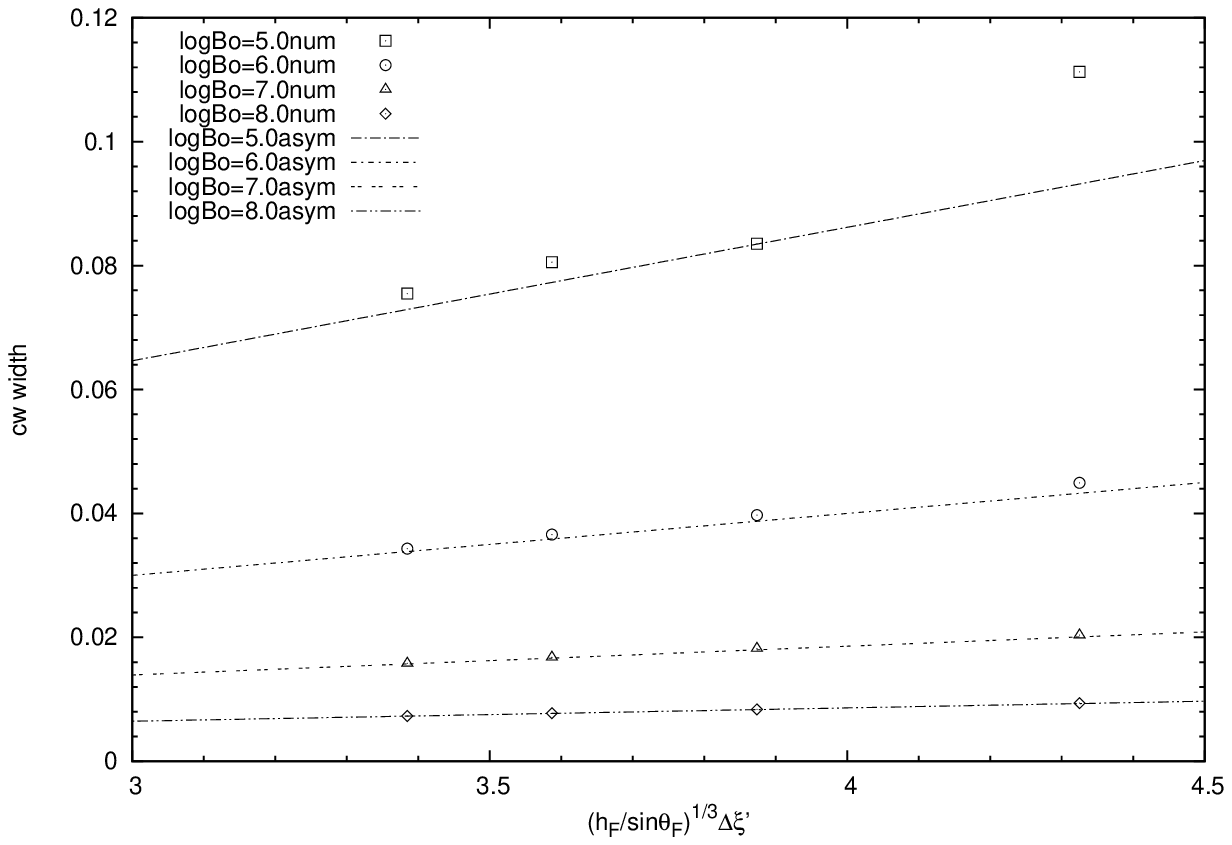}
  \includegraphics[width=0.49\textwidth]{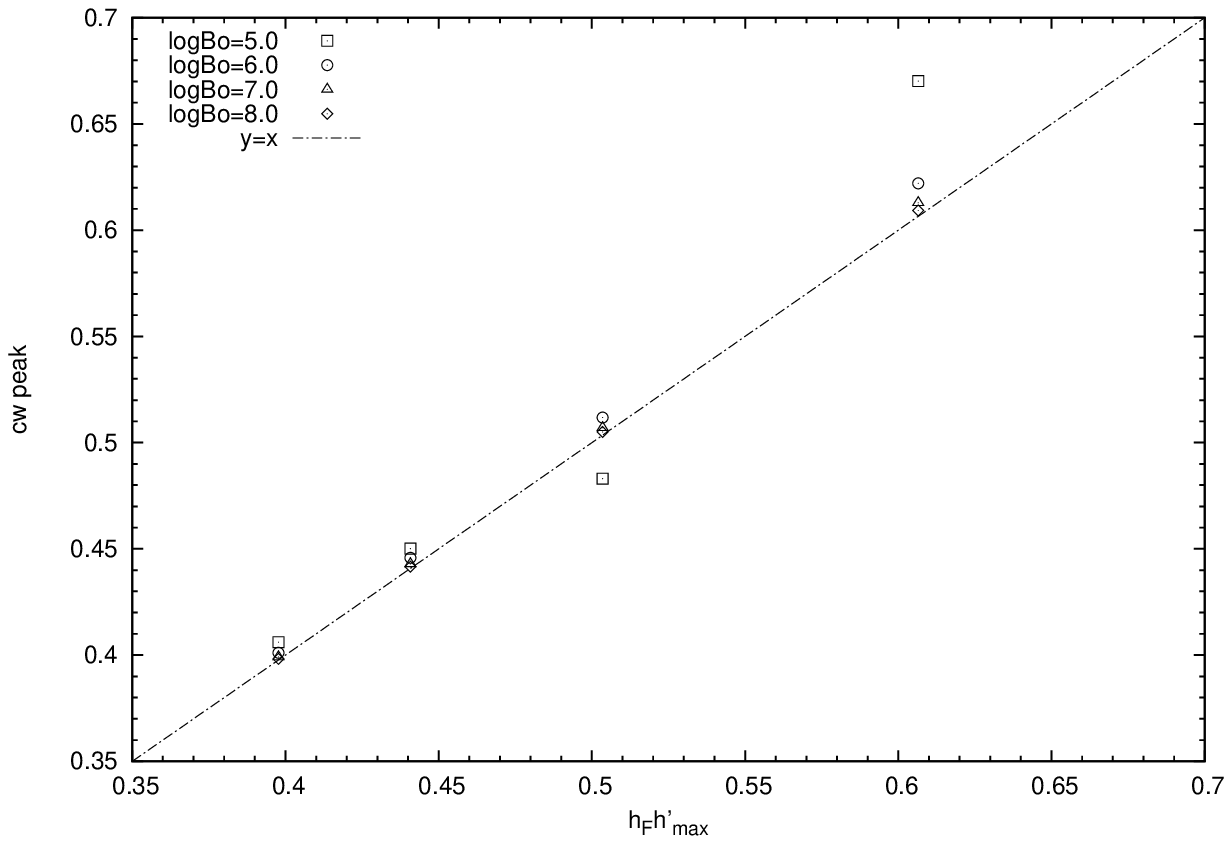}
  \centerline{(a)\hspace{0.45\textwidth}(b)}
  \caption{\label{fig:cywidthpeak} The asymptotic behavior for (a) the width and (b) the peak of capillary waves on cylindrical surface. The precursor thickness is $b=0.001$, the range of $logBo$ is from 5.0 to 8.0 and the range of time is from $t=4$ to $t=10$. The suffix `num' in the legend of (a) represents the data points calculated from direct numerical solution, and the suffix `asym' represents the reference lines generated using the asymptotic theory.}
\end{figure}

One interesting question is that what is the difference between the free-surface profile integrated from inner equation (\ref{eq:innode}) and the profile calculated directly using evolution equation (\ref{eq:tdeq}). This discrepancy as a function of $Bo$ may represent an asymptotic behavior of the capillary ridge, and may be important to study the effect of curved substrate on the complete capillary wave. To describe the global features of the capillary ridge in the inner region, we should define two important parameters - the width and the peak of the capillary wave. We can use these two parameters to quantitatively describe both the inner solutions of (\ref{eq:innode}) and the direct solutions of (\ref{eq:tdeq}). The peak of capillary wave can be defined using the primary maximum value in the inner solution or direct solution. The definition of the width is more complicated, we can use a kind of angular difference to define the width of capillary wave in the direct solution
\begin{equation*}
\Delta\theta=\theta_{\mathrm{pmin}}-\theta_{\mathrm{smin}}
\end{equation*}
where $\theta_{\mathrm{pmin}}$ is the angular position of primary minimum which is described in Fig.~\ref{fig:cycompvst}(b), and $\theta_{\mathrm{smin}}$ is the angular position of secondary minimum which is always located behind the wave peak. We can also use a difference of inner coordinates to define the width of capillary wave in the inner solution
\begin{equation*}
\Delta\xi'=\xi'_{\mathrm{pmin}}-\xi'_{\mathrm{smin}}
\end{equation*}
where $\xi'_{\mathrm{pmin}}$ and $\xi'_{\mathrm{smin}}$ are the inner coordinates of the primary and secondary minimum respectively, as shown in Fig.~\ref{fig:cyoutinnvsb}(b) . Figure~\ref{fig:cywidthpeak} shows the distribution of data points which represent the width of capillary wave under conditions of different Bond number and time. Note that the ordinate is the width captured from the direct solutions. The abscissa is $(\frac{h_{\uF}}{\sin\theta_{\uF}})^{1/3}\Delta\xi'$, where $h_{\uF}$ and $\theta_{\uF}$ are calculated from the corresponding outer solution, and $\Delta\xi'$ is the width captured from the inner solution. The reference lines are calculated according to (\ref{eq:asymw}) and the slope of them is $Bo^{-1/3}$. We can see that the CW (capillary wave) width deviates from the reference line observably when $logBo=5.0$ at early time (with greater $h_{\uF}$ and smaller $\theta_{\uF}$). These deviations are reduced at later time as the front thickness $h_{\uF}$ decreases (front location $\theta_{\uF}$ increases). When $logBo$ is greater than 6.0, the deviation from the reference line is slight at all stage as shown in Fig.~\ref{fig:cywidthpeak}(a). The 1/3-power law (\ref{eq:asymw}) is accurately recovered when $logBo=8.0$. The distribution of the peak of capillary waves for different Bond number and time is shown in Fig.~\ref{fig:cywidthpeak}(b). Note that the ordinate is the peak captured from the direct solutions, and the two factors in abscissa $h_{\uF}h'_{\mathrm{max}}$ are calculated according to (\ref{eq:innerh}) from the corresponding outer and inner solution, respectively. The solid line $y=x$ is generated for reference. The CW peak deviate from the reference line observably when $logBo=5.0$, and present a non-trivial feature (distribute at both side of the reference line) as the front thickness $h_{\uF}$ decreases. But as the Bond number increases, the asymptotic behavior for the peak fits to the $y=x$ line. When $logBo=8.0$, the CW peak recorded from direct solutions are almost identical to the corresponding value $h_{\uF}h'_{\mathrm{max}}$ calculated using the asymptotic theory at all stages, as shown in Fig.~\ref{fig:cywidthpeak}(b).

\begin{figure}
  \centerline{\includegraphics[width=0.8\textwidth]{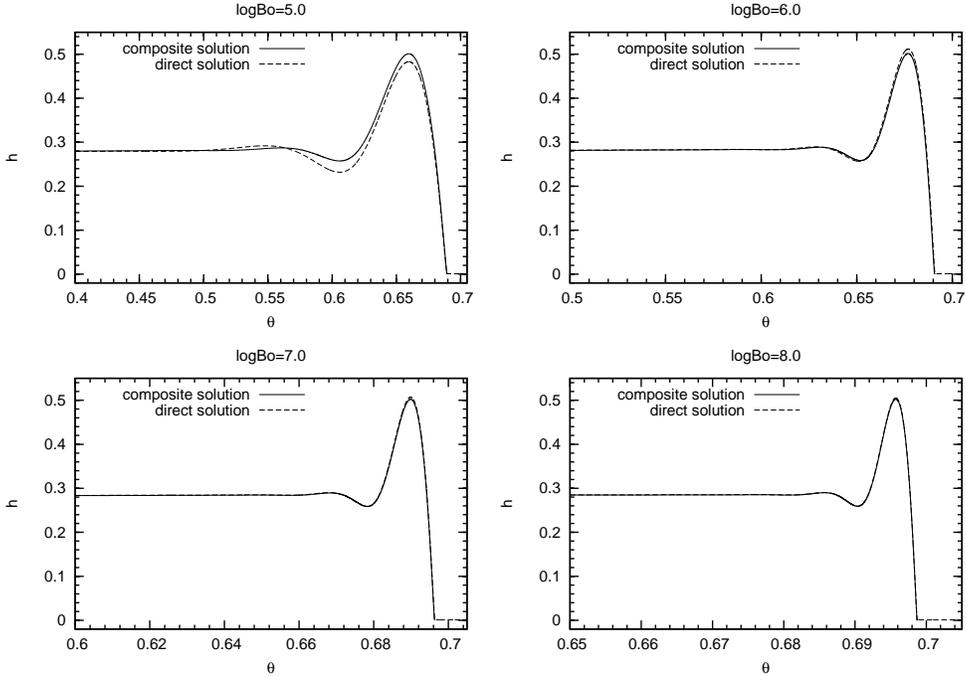}}
  \caption{\label{fig:cyasymbeh} The profile comparisons between composite solutions and direct solutions on cylindrical surface as the Bond number increases. The precursor thickness is $b=0.001$ and the time is at $t=6$.}
\end{figure}

Figure~\ref{fig:cyasymbeh} shows profile comparisons between composite solutions and direct solutions at time $t=6$ as the Bond number increases from $logBo=5.0$ to $logBo=8.0$. The precursor thickness is $b=0.001$ in all examples. Under this precursor thickness, $\delta=b/h_{\uF}$ in (\ref{eq:innode}) is equal to 0.0035 according to the corresponding front thickness at $t=6$. The outer profile obtained from (\ref{eq:cyout}) agrees well with the complete profiles in the outer region, as shown in Fig.~\ref{fig:cyasymbeh}. In the inner region, the asymptotic behavior of capillary ridge is displayed. The width and peak of capillary ridge in the composite profile is slightly different from that in the complete profile when the Bond number is $logBo=5.0$ and $logBo=6.0$. As the Bond number increases, the results show good agreement between the composite and direct solutions. The inner profile in composite solution almost coincides with the capillary ridge in direct solution when $logBo=8.0$. The asymptotic behavior of local features is also clearly demonstrated under high Bond number condition.

\begin{figure}
  \includegraphics[width=0.49\textwidth]{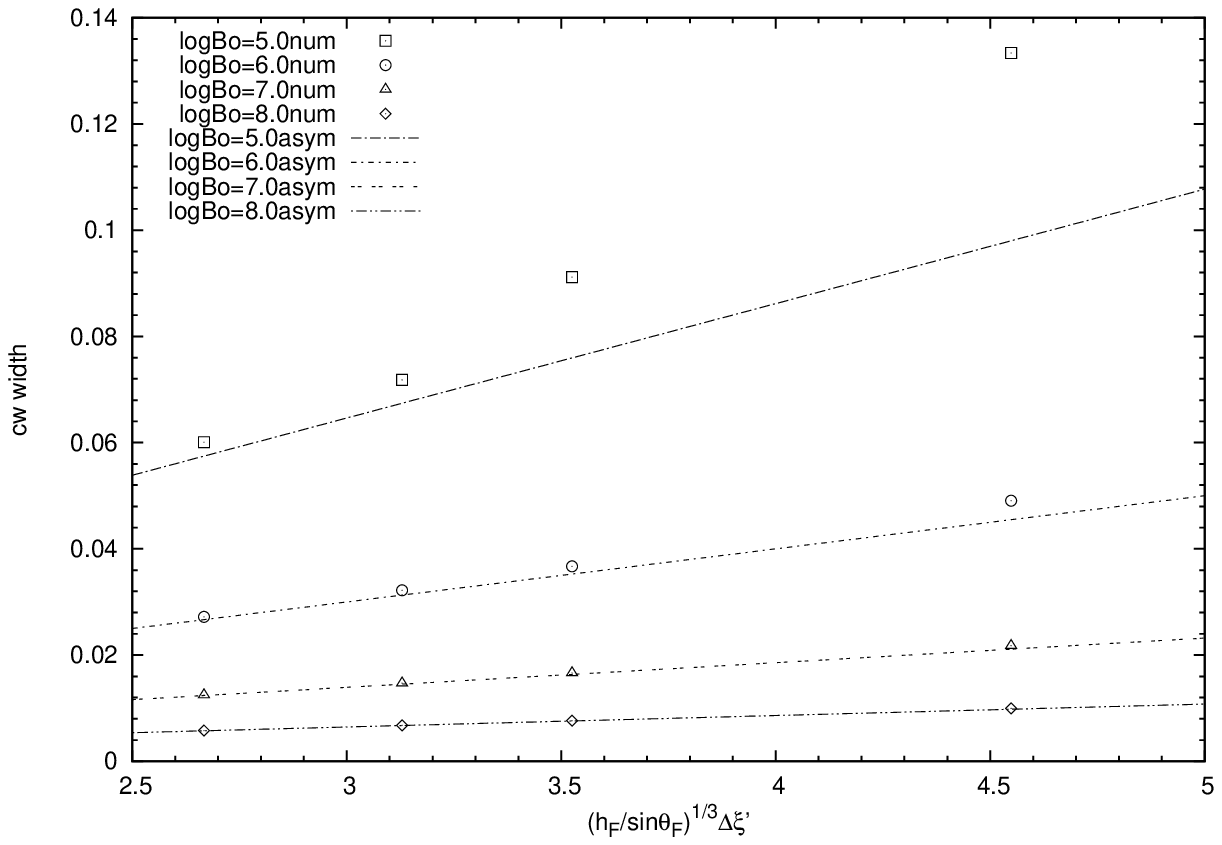}
  \includegraphics[width=0.49\textwidth]{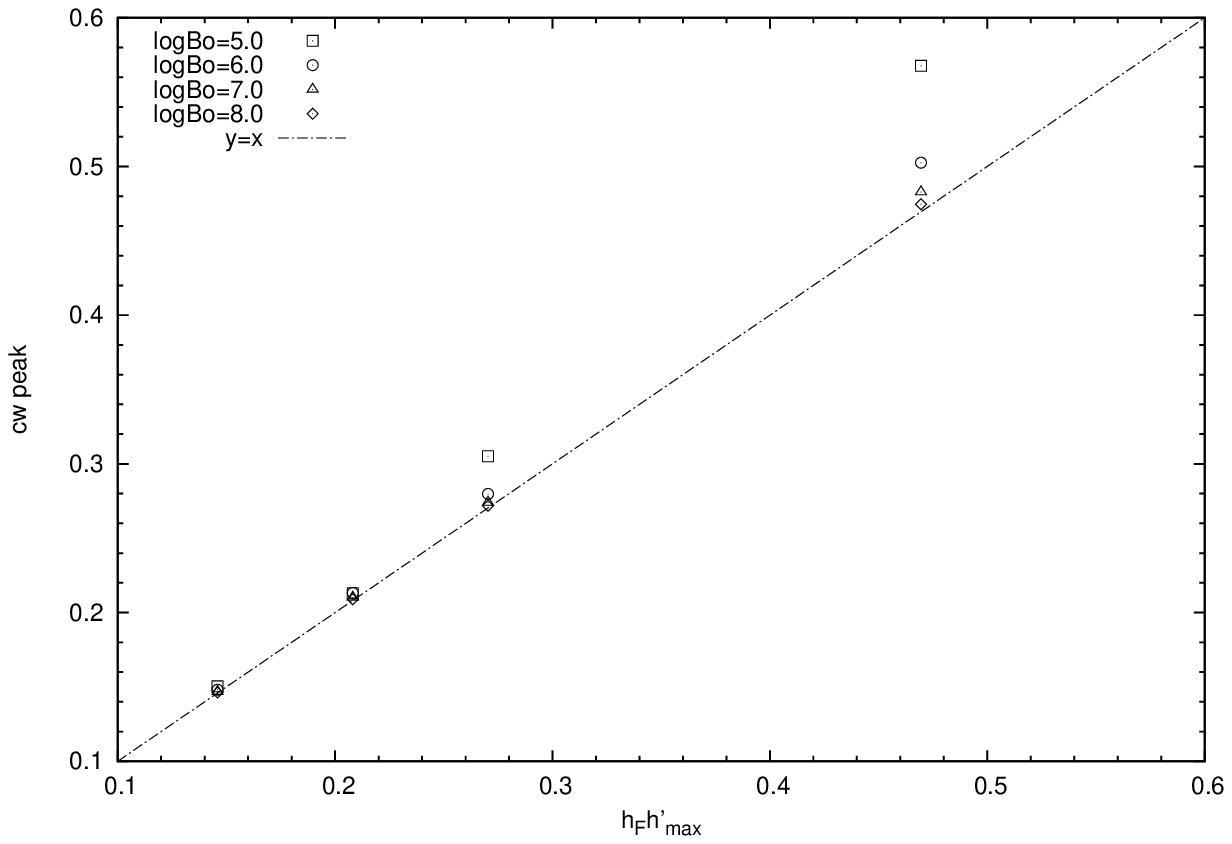}
  \centerline{(a)\hspace{0.45\textwidth}(b)}
  \caption{\label{fig:spwidthpeak} The asymptotic behavior for (a) the width and (b) the peak of capillary waves on spherical surface. The precursor thickness is $b=0.001$, the range of $logBo$ is from 5.0 to 8.0 and the range of time is from $t=10/3$ to $t=100/3$.}
\end{figure}

\begin{figure}
  \centerline{\includegraphics[width=0.8\textwidth]{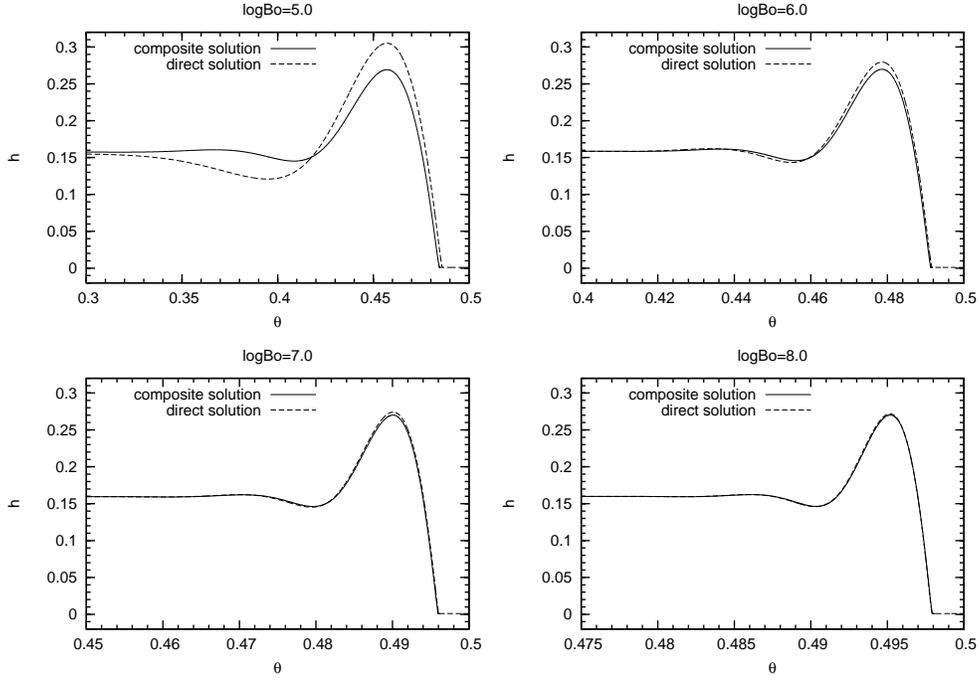}}
  \caption{\label{fig:spasymbeh} The profile comparisons between composite solutions and direct solutions on spherical surface as the Bond number increases. The precursor thickness is $b=0.001$ and the time is at $t=10$.}
\end{figure}

For spherical problem, we do not present the detailed profiles of composite solutions and direct solutions depending on time and Bond number, because there are many similarities in the CW profiles between cylindrical and spherical problem. We only present the asymptotic behavior of the global features using the width and the peak of the capillary wave which have the same definition as the cylindrical problem, as shown in Fig.~\ref{fig:spwidthpeak}(a) and \ref{fig:spwidthpeak}(b). Similar to the cylindrical problem, the discrete data points calculated from (\ref{eq:axeq}) finally fit to the reference lines as the Bond number increases, which indicate the clear asymptotic behavior. The main difference between Fig.~\ref{fig:spwidthpeak} and Fig.~\ref{fig:cywidthpeak} is that for spherical problem the width and peak of the capillary waves deviate more observably from the asymptotic theory when $logBo=5.0$ and $logBo=6.0$. There is much difference between the composite solution and the direct solution at relatively low Bond number. It can be attributed to that there are more first order $O(Bo^{-1/3})$ terms in the complete inner equation for spherical problem (see (\ref{eq:spinnful})) compared to the cylindrical problem (see (\ref{eq:cyinnful})), which makes the leading order inner profiles less accurate. As a contrast to the cylindrical problem, the comparison of profiles between composite solutions and direct solutions on spherical surface at $t=10$ is shown in Fig.~\ref{fig:spasymbeh}. It should be noted that the difference of the profile between the composite solution and direct solution is more obvious than that in cylindrical problem when $logBo=5.0$, which accords with the behavior of the global features. The asymptotic tendency is finally achieved when $logBo$ is greater than 6.0. At $logBo=8.0$, the good agreement of local features between the composite and direct solutions is also obtained.

\subsection{The capillary waves in partial wetting cases}

Now we focus on the partial wetting cases in which the disjoining pressure is operative. The evolution equations (\ref{eq:tddjpeq}) and (\ref{eq:axdjpeq}) are used as the starting points. We have known that there is no disjoining pressure term in the outer equations for both the cylindrical and spherical problems. The profiles of outer solutions are not affected by the addition of disjoining pressure. In the inner region, ODE (\ref{eq:innode}) is replaced by ODE (\ref{eq:inndjpode}). A similar adaptive shooting method which considers the disjoining pressure is also discussed in Appendix~\ref{adx:exoutinn}.

\begin{figure}
  \includegraphics[width=0.49\textwidth]{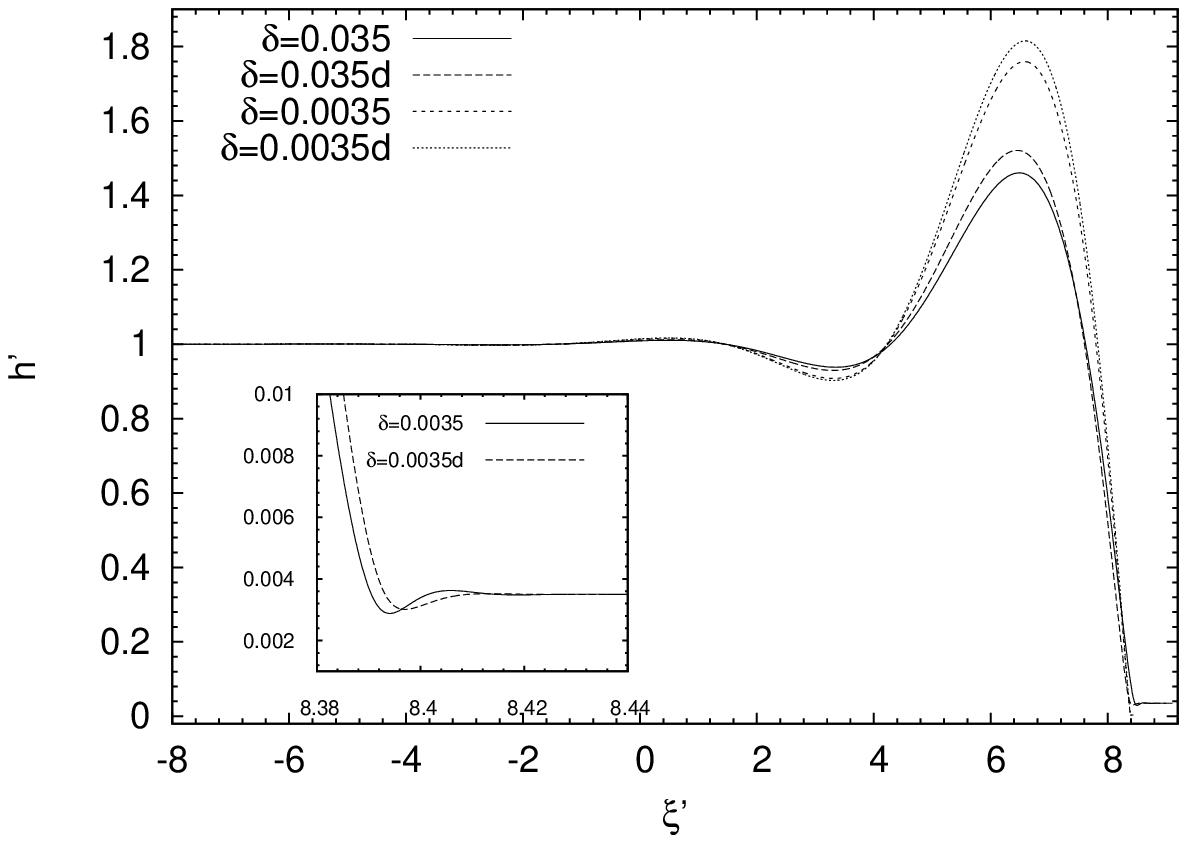}
  \includegraphics[width=0.49\textwidth]{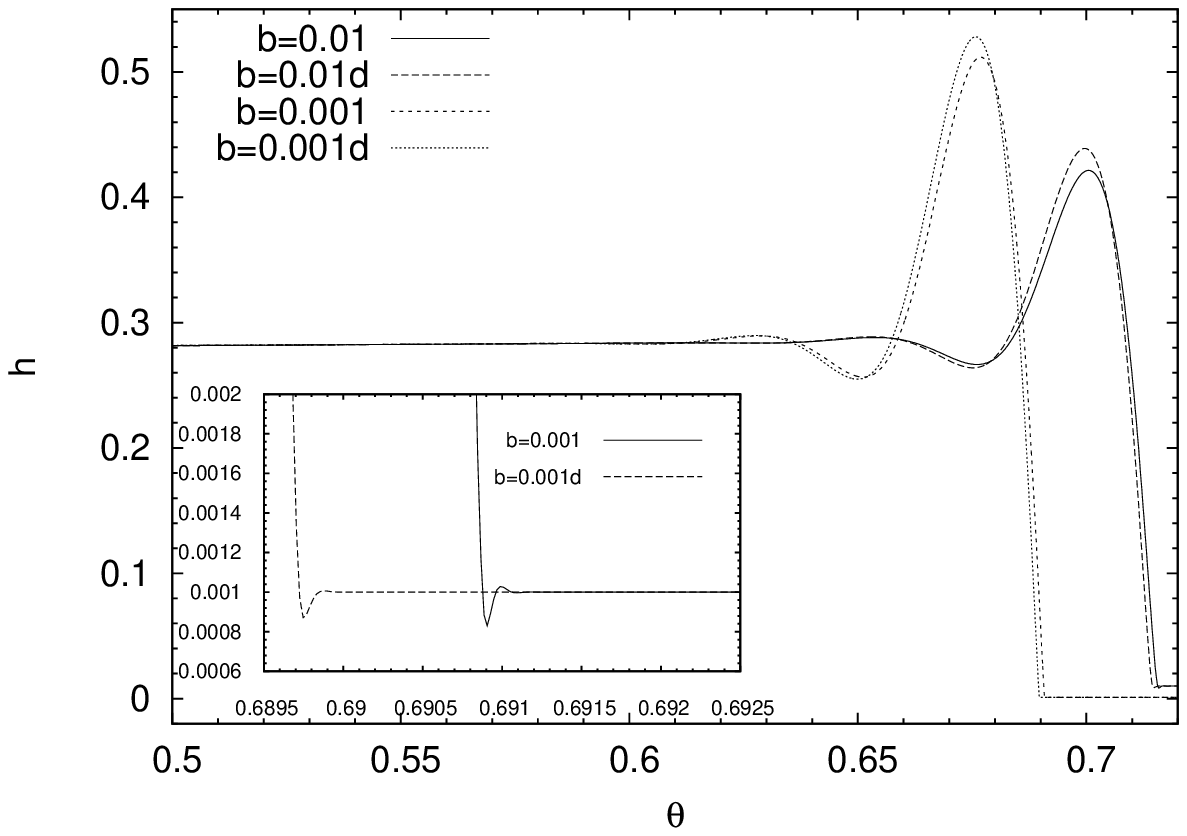}
  \centerline{(a)\hspace{0.45\textwidth}(b)}
  \caption{\label{fig:cydjpcom} Comparison of (a) inner profiles and (b) complete profiles on cylindrical surface in complete and partial wetting cases. The suffix `d' in the legends denotes the disjoining pressure is taken into calculation. The insets in (a) and (b) are the refined structure in contact region. The conditions of precursor thickness are $b=0.01$ and $b=0.001$, and the Bond number in (b) is $logBo=6.0$ at $t=6$. Other parameters are $(n,m)=(3,2)$, $\theta_{\ue}=0.5$ and $\epsilon=0.015$.}
\end{figure}

\begin{figure}
  \includegraphics[width=0.49\textwidth]{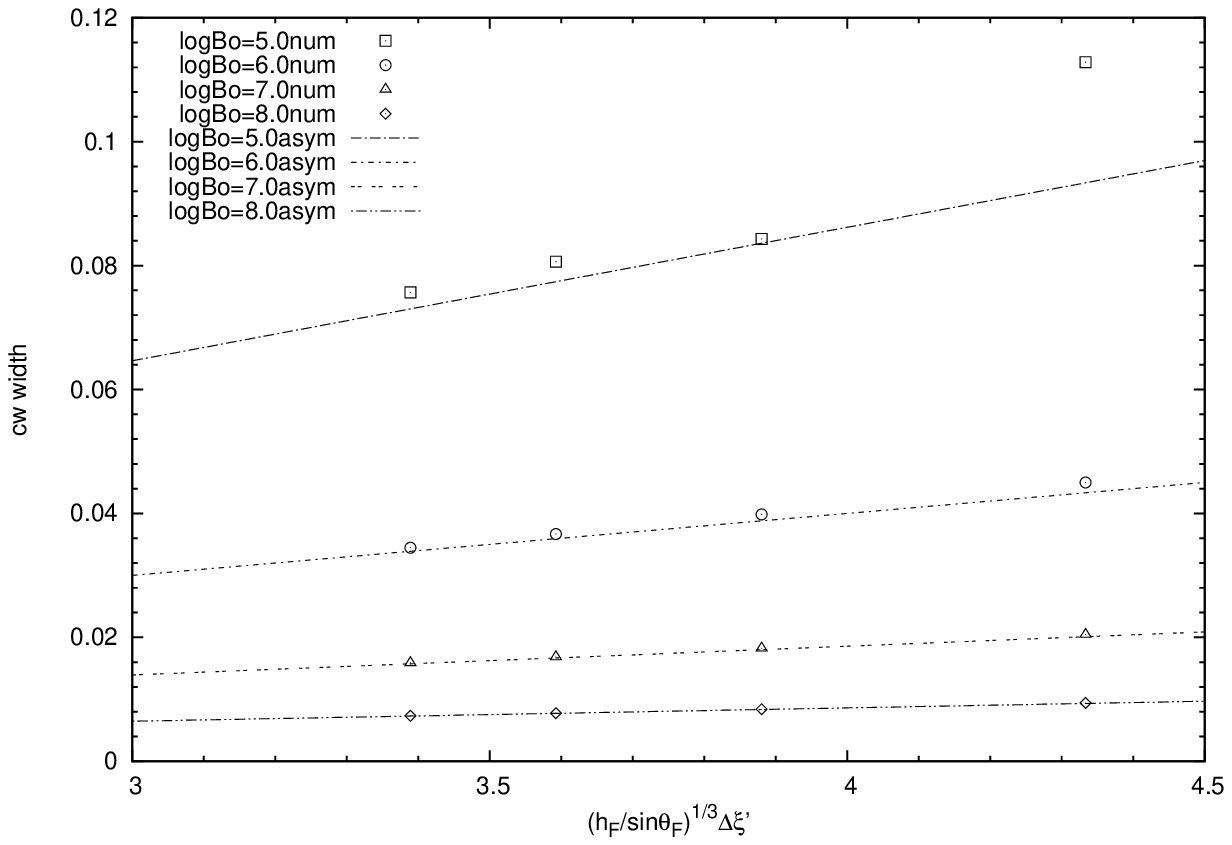}
  \includegraphics[width=0.49\textwidth]{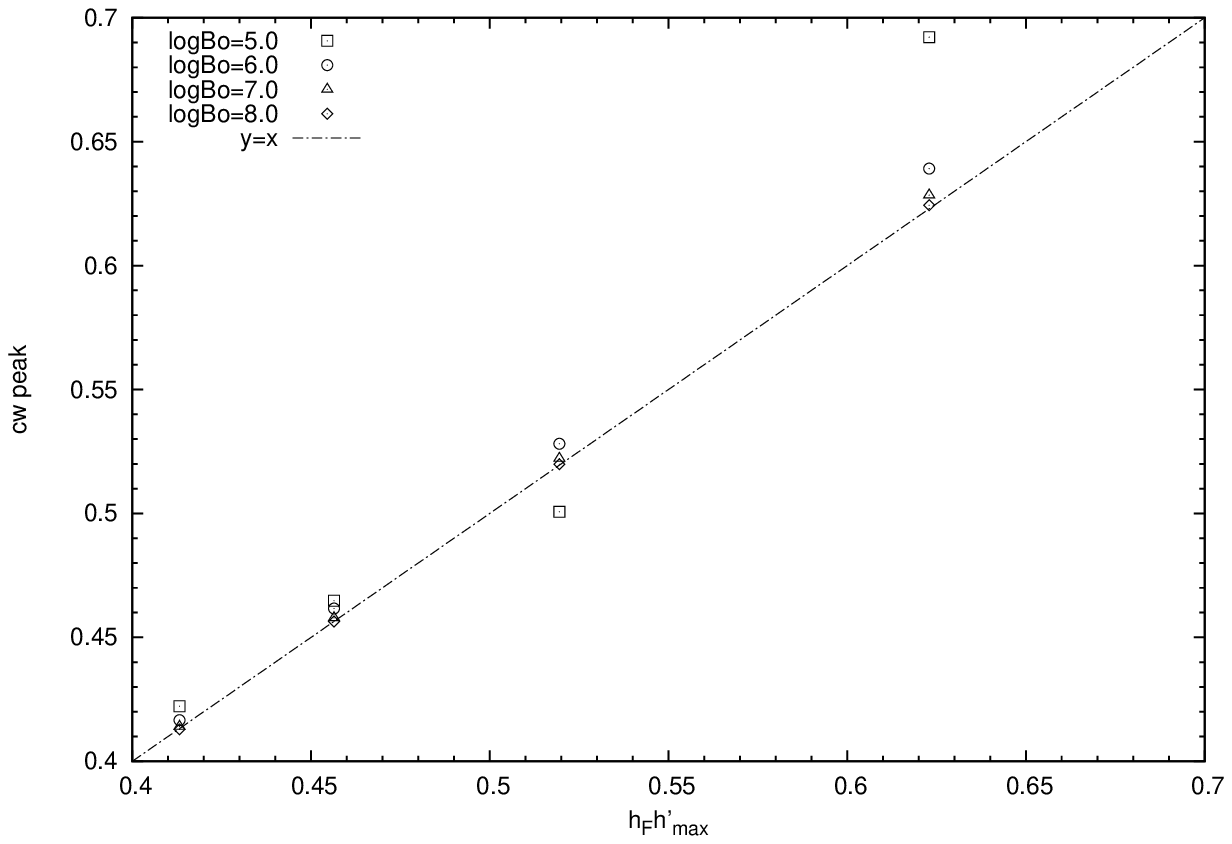}
  \centerline{(a)\hspace{0.45\textwidth}(b)}
  \caption{\label{fig:cydjpwap} The asymptotic behavior for (a) the width and (b) the peak of capillary waves in partial wetting cases on cylindrical surface. The precursor thickness is $b=0.001$, the range of $logBo$ is from 5.0 to 8.0 and the range of time is from $t=4$ to $t=10$. The parameters of disjoining pressure are $(n,m)=(3,2)$, $\theta_{\ue}=0.5$.}
\end{figure}

\begin{figure}
  \centerline{\includegraphics[width=0.8\textwidth]{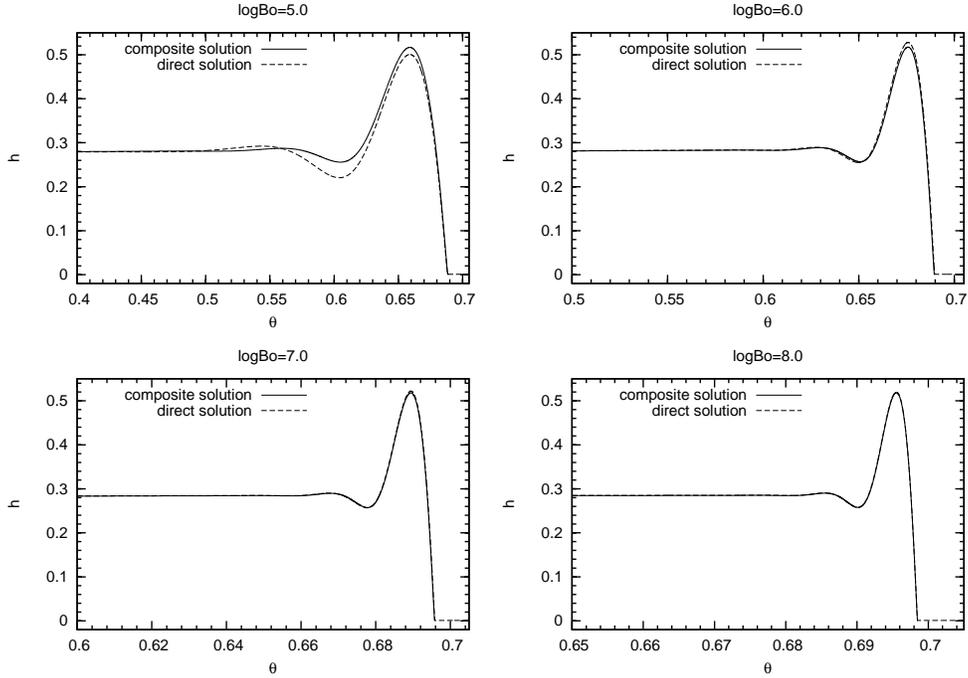}}
  \caption{\label{fig:cyasymbehdjp} The asymptotic behavior under the effect of disjoining pressure on cylindrical surface. The precursor thickness is $b=0.001$ and the time is at $t=6$. The parameters of disjoining pressure are $(n,m)=(3,2)$, $\theta_{\ue}=0.5$.}
\end{figure}

For cylindrical problem, figure~\ref{fig:cydjpcom}(a) shows the comparison between the inner profiles which are integrated from ODE (\ref{eq:innode}) (complete wetting case) and that integrated from ODE (\ref{eq:inndjpode}) (partial wetting case). The relative precursor thickness $\delta=b/h_{\uF}$ is calculated using the front thickness $h_{\uF}$ in outer profiles at $t=6$. Two precursor layer conditions $b=0.01$ and $b=0.001$ are selected for comparative purpose, corresponding to $\delta=0.035$ and $\delta=0.0035$ respectively in ODE (\ref{eq:innode}) and (\ref{eq:inndjpode}). The dimensionless contact angle parameter $K$ in ODE (\ref{eq:inndjpode}) is calculated using (\ref{eq:djpodeco}) with the parameters $(n,m)=(3,2)$, $\theta_{\ue}=0.5$, $logBo=6.0$ and $\epsilon=0.015$. Thus the value of $K$ is 19.57 for $b=0.01$ case and 198.3 for $b=0.001$ case, respectively. There are two main differences for the inner profiles between complete wetting case and partial wetting case. First, the peak and maximum slope of the capillary ridge in partial wetting case is greater than that in complete wetting case. It belongs to the macroscopic effect of the disjoining pressure. Second, the primary minimum near apparent contact line in $\delta=0.0035\ud$ cases is slightly less than that in $\delta=0.0035$ cases, as shown in the inset of Fig.~\ref{fig:cydjpcom}(a), which is the microscopic effect of the disjoining pressure on the refined structure in contact region. The corresponding direct solutions solved from (\ref{eq:tddjpeq}) under the conditions $logBo=6.0$, $t=6$ and $b=0.01$ or $b=0.001$ are shown in Fig.~\ref{fig:cydjpcom}(b). The similar differences like the inner profiles are observed in the inner region, except that the location of the primary minimum in complete wetting case moves forward than that in partial wetting case, clearly seen from the inset. Figure \ref{fig:cydjpwap}(a) and \ref{fig:cydjpwap}(b) show the asymptotic behavior of the width and the peak of the capillary waves on cylindrical surface in partial wetting cases. The parametric conditions are identical to that in Fig.~\ref{fig:cywidthpeak} and the disjoining pressure is set to the same value in Fig.~\ref{fig:cydjpcom}. Compared to Fig.~\ref{fig:cywidthpeak}, the CW width and peak are slightly greater than that in complete wetting case. Except for this main difference compared to the complete wetting case, according to the distribution in Fig.~\ref{fig:cydjpwap}, the global asymptotic behavior is not affected by the addition of disjoining pressure. Under the same parametric conditions ($t=6$ and $b=0.001$) refer to Fig.~\ref{fig:cyasymbeh}, the profiles of the capillary ridges in partial wetting cases are also demonstrated as $logBo$ increases from 5.0 to 8.0, as shown in Fig.~\ref{fig:cyasymbehdjp}. The disjoining pressure does not affect local asymptotic behavior on cylindrical surface.

\begin{figure}
  \includegraphics[width=0.49\textwidth]{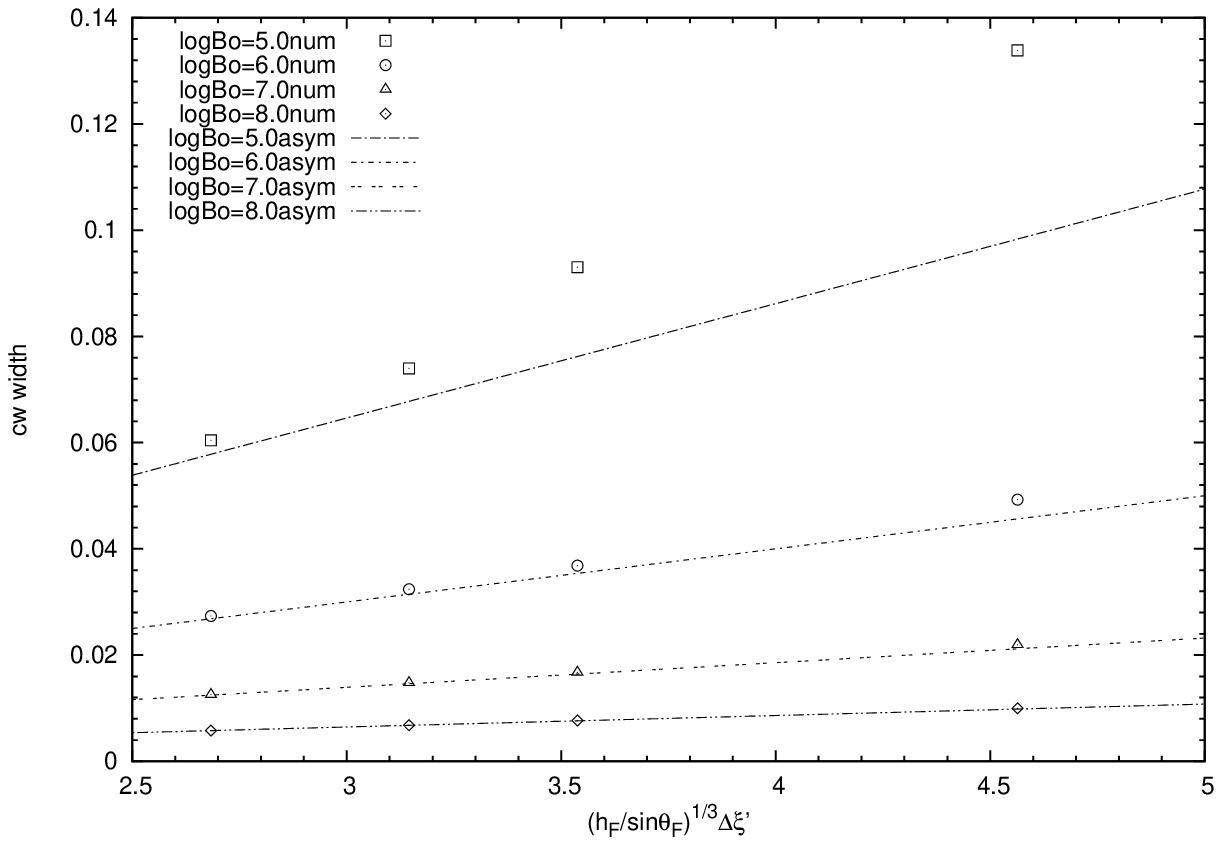}
  \includegraphics[width=0.49\textwidth]{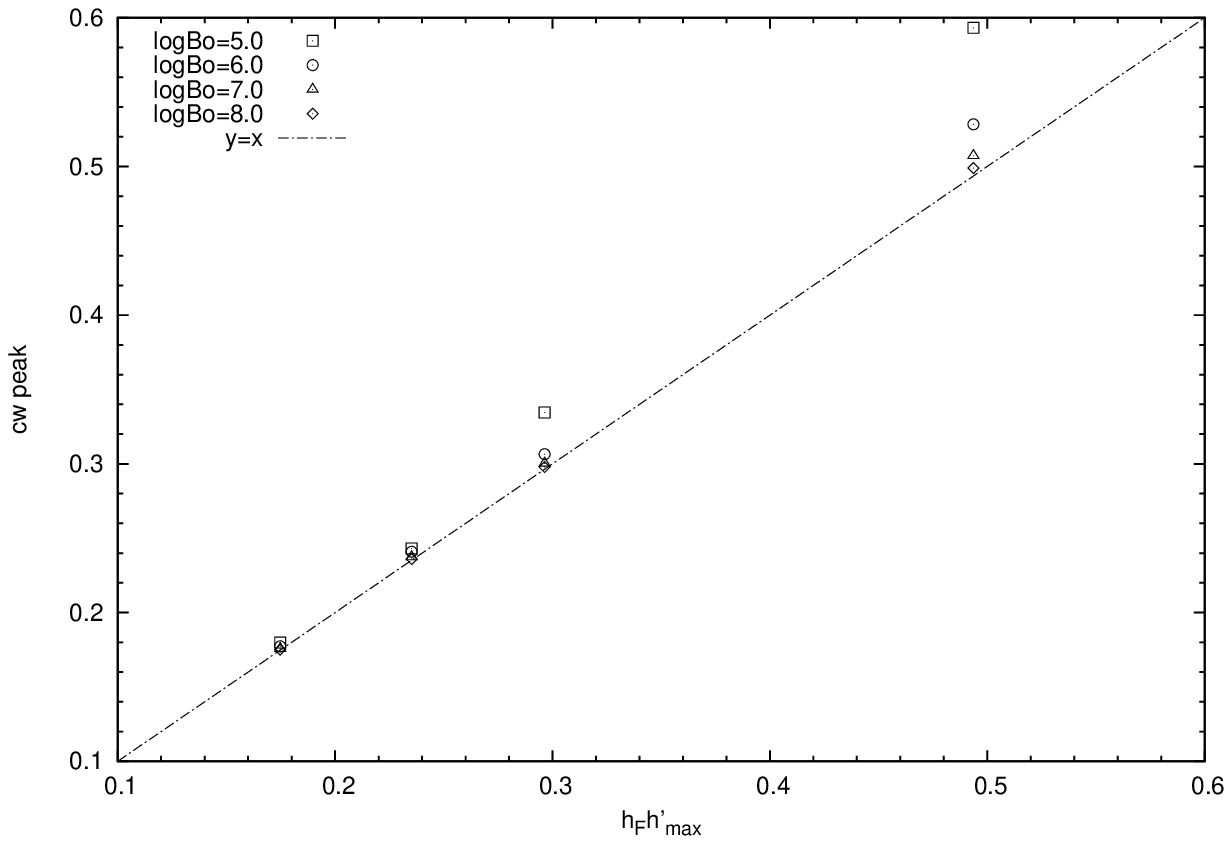}
  \centerline{(a)\hspace{0.45\textwidth}(b)}
  \caption{\label{fig:spdjpwap} The asymptotic behavior for (a) the width and (b) the peak of capillary waves in partial wetting cases on spherical surface. The precursor thickness is $b=0.001$, the range of $logBo$ is from 5.0 to 8.0 and the range of time is from $t=10/3$ to $t=100/3$. The parameters of disjoining pressure are $(n,m)=(3,2)$, $\theta_{\ue}=0.5$.}
\end{figure}

\begin{figure}
  \centerline{\includegraphics[width=0.8\textwidth]{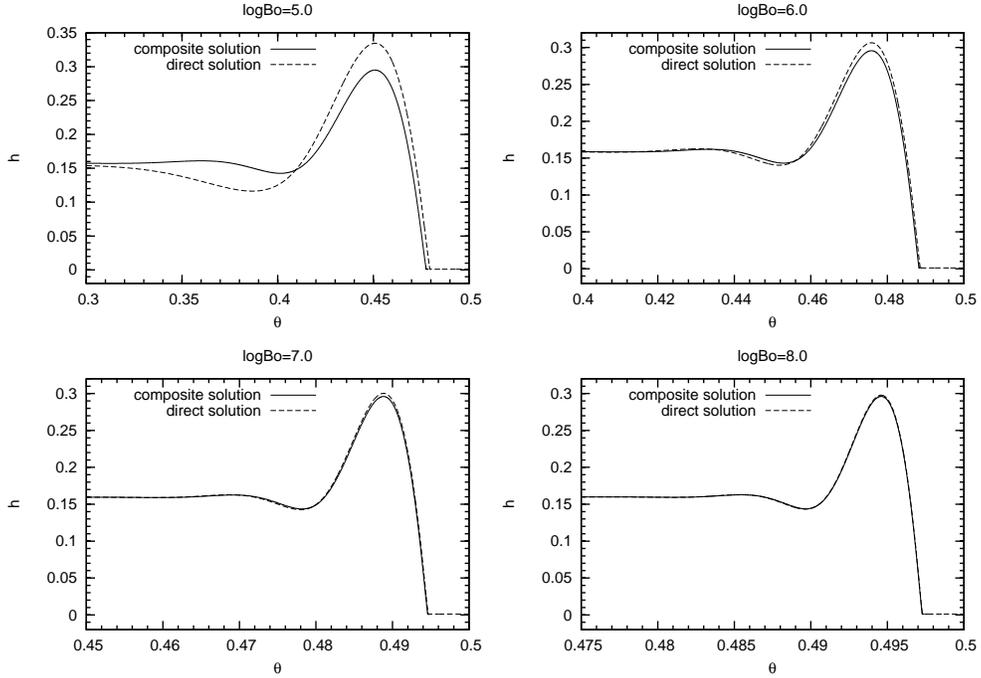}}
  \caption{\label{fig:spasymbehdjp} The asymptotic behavior under the effect of disjoining pressure on spherical surface. The precursor thickness is $b=0.001$ and the time is at $t=10$. The parameters of disjoining pressure are $(n,m)=(3,2)$, $\theta_{\ue}=0.5$.}
\end{figure}

On spherical surface, the comparisons of the CW width and CW peak between the direct numerical results and asymptotic lines are shown in Fig.~\ref{fig:spdjpwap}(a) and \ref{fig:spdjpwap}(b) respectively. Compared to the complete wetting cases as shown in Fig.~\ref{fig:spwidthpeak}, the differences of the CW width and peak induced by disjoining pressure on spherical surface are more observable than that in cylindrical problem (compare Fig.~\ref{fig:cydjpwap} with Fig.~\ref{fig:cywidthpeak}), which may be attributed to the common greater value of $K$ calculated in spherical problems. Figure~\ref{fig:spasymbehdjp} illustrates the asymptotic tendency of the composite profiles under the parametric conditions $t=10$ and $b=0.001$ on spherical surface, $logBo$ is ranging from 5.0 to 8.0 too. According to the plots shown in Fig.~\ref{fig:spdjpwap} and Fig.~\ref{fig:spasymbehdjp}, both of the global and local features of the capillary waves indicate that the asymptotic theory in partial wetting cases is clearly validated for the spherical problem.

\section{\label{sec:conrem}Concluding remarks}

The gravity driven coating flow featuring moving contact line on upper cylinder and hemisphere is formulated using a regular perturbation method which is similar to the lubrication theory for coating problem on inclined plane. The derived leading order governing equations on cylindrical and spherical surfaces depend on one dimensionless Bond number $Bo$, which is determined by the property of the fluid, gravity acceleration, radius of the cylinder or sphere and initial film thickness. Due to the symmetries on cylindrical and spherical surfaces, the governing equations can degenerate into more simplified forms to model two-dimensional flow on cylinder and axisymmetric flow on sphere. With appropriate initial and boundary conditions, a capillary wave solution may be solved from the simplified two-dimensional or axisymmetric evolution equation. A precursor film condition is used to overcome the contact line singularity. In prewetting cases, the precursor thickness is macroscopic. In complete wetting cases, a microscopic precursor film is introduced on the uncoated solid surface. In partial wetting cases, a disjoining pressure model which is compatible with the precursor film is used to simulate the intermolecular forces near the apparent contact line. The effect of disjoining pressure appears as additional terms in the evolution equations.

We emphasize the difference between the high Bond number flow and the low Bond number flow. The high $Bo$ coating flow is focused on in this paper which can be considered as situations that relatively thinner liquid film flow on larger sized solid surface. An asymptotic theory is elaborated to study the high $Bo$ limit of the evolution equations. The method of matched asymptotic expansions is used to derive the outer and inner equations in the outer and the inner region, respectively. Two similar outer equations which only differ in one coefficient are obtained on the cylindrical and spherical surfaces respectively. Because the exact solutions of the primary outer equations may introduce elliptic integrals, we study the leading order outer equations in a top region where the angular position is small. The elementary function solutions can be derived in this small angle region, which can give some analytical features such as the spatial distribution of the free-surface profile and the long-term scaling law of the moving front location. As the most important deduction of this asymptotic theory, the leading order inner equations on cylindrical surface and spherical surface are identical, and the effect of curved substrate can be considered as higher order corrections. It implies that the capillary waves on cylindrical surface and spherical surface may have inherent similarities in the inner region. A 1/3-power law is obtained for the width of capillary ridge on both the cylindrical and spherical surfaces. The width is proportional to the cubic root of front thickness, and inversely proportional to the cubic root of Bond number and sine of front location. On the other hand, the peak of capillary ridge can be calculated by multiplying the front thickness and the peak value of the inner solution. The matched method for constructing a composite capillary wave in the whole flow domain is given by combining the profiles in the outer region and inner region. The asymptotic theory in partial wetting cases is also discussed. The disjoining pressure term will not appear in the outer equations, but is added on the right side of the inner equation using a common form for both the cylindrical and spherical problems.

Appropriate numerical methods are proposed for solving the outer equations, inner equations and the two-dimensional and axisymmetric evolution equations. The exact solutions of the outer and inner equations are demonstrated. There are some common features between the outer solutions on cylindrical surface and that on spherical surface. The profile is homogeneous at small angle region, but monotonically increases as the angle increases. The long-term spreading can be described using different scaling laws so that the spreading speed of the moving front is essentially different. In the inner system, the profiles of fluctuant capillary ridge have similarity, which can be attributed to the identical inner equation in the sense of leading order for both problems. The moving front speed can be approximated using the propagation speed of the shock wave in the outer solutions. The exact front location as a function of time accords with the long-term scaling laws at early time, but deviates from the scaling laws obviously at later time, because the prerequisite of these scaling laws is that the angular position should be sufficiently small. Using the matched method given in the asymptotic theory, the composite profiles are constructed and compared to the direct solutions of the evolution equations. The asymptotic behavior in the inner region is shown via both the global and local features. Two parameters including the width and peak of the capillary waves are introduced for describing the global features. The width and peak in composite solutions are found to be closer to the direct solutions as the Bond number increases. The local features are shown by profile comparisons between composite solutions and direct solutions as $Bo$ increases and the agreement of local profiles is also obtained for high $Bo$ cases. This asymptotic behavior is common for both cylindrical and spherical problems and provides powerful numerical proofs for the asymptotic theory. The effect of disjoining pressure is demonstrated by comparing the inner profiles and complete profiles in complete wetting case to that in partial wetting case. The global and local asymptotic behavior on cylindrical and spherical surfaces is not affected by the addition of disjoining pressure.

The asymptotic theory in present work provides a theoretical method to construct exact capillary waves for high Bond number coating flow on cylinder and sphere (if the iterative solution of a nonlinear algebraic equation and the numerical integration of an ODE can be loosely called ``exact'' results). This theory can be extended to study the coating flow on other two-dimensional or axisymmetric solid surface such as parabolic cylinder or spheroid. The method of matched asymptotic expansions used here gives an idea to deal with high Bond number coating flow on a general curved surface. The profile of capillary ridge on inclined plane is universal for gravity driven capillary wave on arbitrary surface because a sufficiently small inner region on curved surface can be regarded as a planar region. It is believed that the asymptotic theory is invalid in a low Bond number flow (thicker film flow down smaller sized surface), but the evolution equations still hold to the leading order. For this kind of flow, the numerical approach is an effective way and the high resolution adaptive method proposed here can be used to obtain numerical capillary waves. It may be the future focus for the research of coating flow on curved surface. Moreover, as described in Sec.~\ref{sec:intro}, the quantitative profile of the capillary wave is the basis for the study of fingering instability on cylinder and sphere, which we will describe in a future paper.

\appendix

\section{\label{adx:exoutinn}Exact solutions for outer and inner equations}

The exact solutions of the outer equations (\ref{eq:cyout}) and (\ref{eq:spout}) can be obtained in implicit form via the method of characteristics. For cylindrical problem, consider the curve of characteristics
\begin{equation}\label{eq:cychl}
\frac{\ud\theta}{\ud t}=3h^{2}[\theta(t),t]\sin\theta
\end{equation}
Upon this curve (\ref{eq:cyout}) degenerates into an ordinary differential equation
\begin{equation}\label{eq:cyhode}
\frac{\ud h}{\ud t} =-h^{3}\cos\theta
\end{equation}
Equations (\ref{eq:cychl}) and (\ref{eq:cyhode}) are solved subject to initial conditions $\theta(0)=\theta_0$ and $h(\theta,0)=h_{\ui}(\theta)$, which leads to the following implicit form of the solution
\begin{subequations}\label{eq:cyoutsol}
\begin{equation}\label{eq:cyht}
h(\theta,t)=\left\{\begin{array}{ll}
                     (2t+\frac{1}{h_{\ui}^{2}(\theta)})^{-\frac{1}{2}}  & (\theta=\theta_0=0) \\
                     h_{\ui}(\theta _0)(\frac{\sin\theta_0}{\sin\theta})^{\frac{1}{3}} & (0<\theta<\pi)
                 \end{array}\right.
\end{equation}
in which the first expression at $\theta=0$ is identical to the general solution (\ref{eq:cyoutgsol}) in small $\theta$ region. The time dependence of the azimuthal angle is given implicitly as the solution of
\[\mathrm{W}_{1}(\sin^{\frac{2}{3}}\theta)-\mathrm{W}_{1}(\sin^{\frac{2}{3}}\theta_0)=3h_{\ui}^{2}(\theta_0)(\sin\theta_0)^{\frac{2}{3}}t\]
where $\mathrm{W}_1(x)$ is the first kind elliptic integral of Weierstrass's form. For convenience of computation, the Weierstrass's form should be transformed into the Legendre's form \citep{Erdelyi1953}, the above equation can be modified as (for $0<\theta\leq\pi/2$)
\begin{equation}\label{eq:cythetat}
F[g(\theta),\sin(\frac{5\pi}{12})]-F[g(\theta_0),\sin(\frac{5\pi}{12})]+2(3)^{\frac{1}{4}}h_{\ui}^{2}(\theta_0)\sin^{\frac{2}{3}}\theta_{0}t=0
\end{equation}
where
\begin{equation}\label{eq:fellipi}
F(\phi,k)=\int_{0}^{\phi}\frac{\ud\vartheta}{\sqrt{1-k^2\sin^2\vartheta}}
\end{equation}
is the first kind incomplete elliptic integral of Legendre's form whose first argument is
\begin{equation}\label{eq:funcg}
g(\vartheta)=\arccos(\frac{\sqrt{3}-1+\sin^{\frac{2}{3}}\vartheta}{\sqrt{3}+1-\sin^{\frac{2}{3}}\vartheta})
\end{equation}
\end{subequations}
This form is identical to a large Bond number case shown in \citet{Reisfeld1992}, in which an implicit solution for a liquid film flow on a horizontal cylinder is presented via almost the same expression of the first kind incomplete elliptic integral, even though the initial film considered there is assumed to be uniformly distributed on both the upper and lower cylinder.

For a given time $t$ and a given location $\theta$, the parameter $\theta_0$ can be numerically solved from nonlinear algebraic equation (\ref{eq:cythetat}) using a Newton's method, then the film thickness $h$ is calculated using (\ref{eq:cyht}). To construct a complete outer profile at a certain time, a number of (\ref{eq:cythetat}) with different values of $\theta$ and a same value of $t$ constitute a system of independent algebraic equations for obtaining the film thickness at different locations. Note that there is no difficulty for the Newton's method to solve (\ref{eq:cythetat}), because the value of incomplete elliptic integral can be evaluated using the standard subroutine in module of special functions and the derivative is analytical. If the outer equation (\ref{eq:cyout}) is solved subject to initial condition (\ref{eq:cwic1}), the value of parameter $\theta_0$ calculated from (\ref{eq:cythetat}) may become multi-valued. It implies that the characteristics lines may intersect due to the discontinuity in initial condition and a shock wave may be formed. The location of shock wave (front location) $\theta_\uF$ can be determined by volume conservation
\begin{equation}\label{eq:cyvolcons}
\int_{0}^{\theta_{\uF}}[h_{1}(\theta,t)-h_{b}(\theta,t)]\ud\theta=V
\end{equation}
where the profiles $h_1$ and $h_b$ are constructed via separately solving two systems of equations (\ref{eq:cyoutsol}) by setting the initial function $h_\ui(\theta_0)=1$ and $h_\ui(\theta_0)=b$, respectively. After the front location is calculated, the front thickness $h_\uF$ is obtained as $h_\uF=h_1(\theta_\uF,t)$.

For spherical problem, a similar procedure of the method of characteristics can be applied. Upon the curve of characteristics
\begin{equation}\label{eq:spchl}
\frac{\ud\theta}{\ud t}=3h^{2}[\theta(t),t]\sin\theta
\end{equation}
(\ref{eq:spout}) reduces to the ordinary differential equation
\begin{equation}\label{eq:sphode}
\frac{\ud h}{\ud t} =-2h^{3}\cos\theta
\end{equation}
subject to initial conditions $\theta(0)=\theta_0$ and $h(\theta,0)=h_{\ui}(\theta)$. The solution can be also expressed using an implicit form
\begin{subequations}\label{eq:spoutsol}
\begin{equation}\label{eq:spht}
h(\theta,t)=\left\{\begin{array}{ll}
                     \frac{1}{2}(t+\frac{1}{4h_{\ui}^{2}(\theta)})^{-\frac{1}{2}}  & (\theta=\theta_0=0) \\
                     h_{\ui}(\theta _0)(\frac{\sin\theta_0}{\sin\theta})^{\frac{2}{3}} & (0<\theta<\pi)
                 \end{array}\right.
\end{equation}
where the first line is identical to the solution (\ref{eq:spoutsol1}) in small $\theta$ region when $h_\ui(0)=1$. The parameter $\theta_0$ as a function of time can be integrated as an implicit form
\[\mathrm{W}_{2}(\sin^{\frac{2}{3}}\theta)-\mathrm{W}_{2}(\sin^{\frac{2}{3}}\theta_0)=3h_{\ui}^{2}(\theta_0)(\sin\theta_0)^{\frac{4}{3}}t\]
where $\mathrm{W}_2(x)$ is the second kind elliptic integral of Weierstrass's form. Note that the second kind elliptic integral of Weierstrass's form can be transformed into the summation of the first kind elliptic integral of Legendre's form, the second kind elliptic integral of Legendre's form and two elementary integrals which have analytical forms. The above equation can be rewritten as (for $0<\theta\leq\pi/2$)
\begin{multline}\label{eq:spthetat}
(\sqrt{3}-1)\{F[g(\theta),\sin(\frac{5\pi}{12})]-F[g(\theta_0),\sin(\frac{5\pi}{12})]\}-2\sqrt{3}\{E[g(\theta),\sin(\frac{5\pi}{12})]-\\
E[g(\theta_0),\sin(\frac{5\pi}{12})]\}+2\sqrt{3}\{D[g(\theta)]-D[g(\theta_0)]\}-2(3)^{\frac{1}{4}}h_{\ui}^{2}(\theta_0)\sin^{\frac{4}{3}}\theta_{0}t=0
\end{multline}
where
\begin{equation}\label{eq:sellipi}
E(\phi,k)=\int_{0}^{\phi}\sqrt{1-k^2\sin^2\vartheta}\ud\vartheta
\end{equation}
is the second kind incomplete elliptic integral of Legendre's form whose first argument is identical to (\ref{eq:funcg}) and
\begin{equation}\label{eq:funcD}
D(\phi)=\frac{(1-\cos\phi)\sqrt{1-k^2\sin^2\phi}}{\sin\phi}
\end{equation}
\end{subequations}
is an elementary function. The Newton's method to solve nonlinear algebraic equation (\ref{eq:spthetat}) and the numerical technique to construct the outer profile is analogous to that in cylindrical problem except that the volume formula for calculating the front location $\theta_\uF$ is different
\begin{equation}\label{eq:spvolcons}
\int_{0}^{\theta_{\uF}}[h_{1}(\theta,t)-h_{b}(\theta,t)]\sin\theta\ud\theta=V
\end{equation}
where the profiles $h_1$ and $h_b$ are constructed by setting the initial function $h_\ui(\theta_0)=1$ and $h_\ui(\theta_0)=b$, respectively.

The boundary value problem in the inner region can be solved using a classic shooting method for a given $\delta$ \citep{Tuck1990}. The third order ODE (\ref{eq:innode}) can be re-written as three first order ODEs, and the initial values should be set at a certain point. Because the boundary value of (\ref{eq:innode}) is located at infinitely far, an asymptotic equation valid far upstream should be used to set the initial conditions. When $\xi'\rightarrow-\infty$, the uniform layer is only slightly perturbed
\[h'\rightarrow 1+g'\]
The linear ODE for perturbation $g'$ is derived from the degeneration of (\ref{eq:innode})
\begin{equation}\label{eq:uplimode}
\frac{\ud^{3}g'}{\ud\xi'^{3}}+(2-\delta-\delta^2)g'=0
\end{equation}
The characteristic equation of (\ref{eq:uplimode}) is
\begin{equation}\label{eq:cheq}
q^3+2-\delta-\delta^2=0
\end{equation}
Thus the upstream limiting solution of (\ref{eq:innode}) is
\begin{equation}\label{eq:uplimsol}
h'\rightarrow 1+ae^{b\xi'}\cos(c\xi')
\end{equation}
where $a$ is an arbitrary parameter, and $b$ and $c$ are the real and imaginary parts of a conjugate complex root $q$ of the characteristic equation (\ref{eq:cheq}). The initial values of the first order ODEs $h'$, $h'_{\xi'}$, and $h'_{\xi'\xi'}$ can be calculated using limiting solution (\ref{eq:uplimsol}) at a sufficiently far location upstream. Then this initial value problem is solved using a fourth-order Runge-Kutta method. The parameter $a$ is a free parameter in the shooting process and can vary until the downstream condition $h'\rightarrow\delta$ is satisfied to high accuracy. Finally the inner profiles can be numerically integrated. The integration step of the Runge-Kutta method should be adaptive because the profile in the inner region and contact region varies sharply, especially for the small $\delta$ case. The integration step can be adapted according to the local values of the first order and second order derivatives of the numerical profile. An alternative adaptive function can be expressed as
\begin{equation}\label{eq:adpfun}
\Delta\xi'=\frac{\Delta\xi'_{0}}{\sqrt{1+\alpha h_{\xi'}^{2}+\beta h_{\xi'\xi'}^{2}}}
\end{equation}
Where $\Delta\xi_0'$ is the uniform integration step, $\alpha$ and $\beta$ are two adjusting factors.

The ODE (\ref{eq:inndjpode}) which includes the disjoining pressure terms can be solved using the similar shooting method described to solve ODE (\ref{eq:innode}). But the characteristic equation of the linear ODE which is valid at far upstream is
\begin{equation}\label{eq:cheqdjp}
q^3+K(m\delta^m-n\delta^n)q+2-\delta-\delta^2=0
\end{equation}
Except that, the shooting steps are identical to that for solving (\ref{eq:innode}).

\section{\label{adx:numtdax}Numerical schemes for evolution equations}

The conservation form of two-dimensional evolution equation (\ref{eq:tdeq}) is
\begin{subequations}
\begin{equation}\label{eq:tdeqc}
\frac{\partial h}{\partial t}+\frac{\partial Q}{\partial\theta}=0
\end{equation}
\begin{equation}\label{eq:tdeqflux}
Q=h^{3}(-\frac{\partial p}{\partial\theta}+\sin\theta)
\end{equation}
\begin{equation}\label{eq:tdeqp}
p=-\frac{1}{Bo}(h+\frac{\partial^{2}h}{\partial\theta^{2}})
\end{equation}
\end{subequations}
To overcome the numerical limitation of an explicit scheme, we use a semi-implicit Crank-Nicolson scheme to discretize (\ref{eq:tdeqc}), (\ref{eq:tdeqflux}) and (\ref{eq:tdeqp})
\begin{subequations}
\begin{equation}\label{eq:numtdc}
\frac{h_{i}^{k+1}-h_{i}^{k}}{\Delta t}+\frac{1}{2}\frac{1}{\Delta\theta}(Q_{i+1/2}^{k+1}-Q_{i-1/2}^{k+1})+
\frac{1}{2}\frac{1}{\Delta\theta}(Q_{i+1/2}^{k}-Q_{i-1/2}^{k})=0
\end{equation}
\begin{equation}\label{eq:numtdflux}
Q_{i+1/2}^{k+1}=[0.5(h_{i}^{k+1}+h_{i+1}^{k+1})]^{3}(-\frac{p_{i+1}^{k+1}-p_{i}^{k+1}}{\Delta\theta}+\sin\theta_{i+1/2})
\end{equation}
\begin{equation}\label{eq:numtdp}
p_{i}^{k+1}=-\frac{1}{Bo}(h_{i}^{k+1}+\frac{h_{i+1}^{k+1}+h_{i-1}^{k+1}-2h_{i}^{k+1}}{\Delta\theta^{2}})
\end{equation}
\end{subequations}
The second order derivative of $h$ in (\ref{eq:tdeqp}) is calculated by the typical central difference scheme shown in (\ref{eq:numtdp}). For a given Bond number $Bo$, a system of quartic equations for the film thickness $h_i^{k+1}$ at the next time step is constructed by substituting (\ref{eq:numtdflux}) and (\ref{eq:numtdp}) into (\ref{eq:numtdc}). We use an iterative Newton-Kantorovich's method to solve the nonlinear equations.

For axisymmetric evolution equation (\ref{eq:axeq}), the conservation form is
\begin{subequations}
\begin{equation}\label{eq:axeqc}
\frac{\partial h}{\partial t}+\frac{1}{\sin\theta}\frac{\partial}{\partial\theta}(\sin\theta Q)=0
\end{equation}
\begin{equation}\label{eq:axeqflux}
Q=h^{3}(-\frac{\partial p}{\partial\theta}+\sin\theta)
\end{equation}
\begin{equation}\label{eq:axeqp}
p=-\frac{1}{Bo}(2h+\frac{1}{\tan\theta}\frac{\partial h}{\partial\theta}+\frac{\partial^{2}h}{\partial\theta^{2}})
\end{equation}
\end{subequations}
A similar Crank-Nicolson scheme is used for (\ref{eq:axeqc}), (\ref{eq:axeqflux}) and (\ref{eq:axeqp}), the difference is that the sinusoidal coefficient in (\ref{eq:axeqc}) and the additional first order derivative of $h$ in (\ref{eq:axeqp}) which can be discretized using the central difference scheme. For evolution equations (\ref{eq:tddjpeq}) and (\ref{eq:axdjpeq}) including the disjoining pressure terms, because the derivatives of $h$ do not appear in the disjoining pressure equation (\ref{eq:djpd}), the discrete form can be written directly
\begin{equation}\label{eq:numdjp}
{\Pi_{\ud}}_{i}^{k}=\frac{1}{\epsilon^{2}}\frac{(n-1)(m-1)}{2b(n-m)}\theta_{\ue}^{2}[(\frac{b}{h_i^k})^{n}-(\frac{b}{h_i^k})^{m}]
\end{equation}

The time spent by this numerical scheme may increase observably with the total number of nodal points in the solved region because of the iterative process at each time step. A common \emph{r}-adaptive method is to use the adaptive function like (\ref{eq:adpfun}) according to the derivative and curvature of the free-surface profile to redistribute the nodes in the whole domain, then to smooth the relocated nodes for avoiding the extra errors from uneven nodal distribution introduced by the previous adaptive step. But this smoothing process may be insufficient to eliminate the extra errors in the problems considered here because the sharp fluctuation in the inner region and contact region may result in large gradient of the grid size which may not be smoothed effectively. We can use some priori knowledge of the capillary wave profile to design a simple and appropriate \emph{r}-adaptive method. Because under high Bond number condition the capillary ridge is located in a small region near the apparent contact line, a smooth hyperbolic tangent distribute function can be used to relocate the nodes
\begin{equation}\label{eq:disfun}
s(i)=1+\frac{\tanh(\frac{\alpha}{2}\frac{i-N}{N-1})}{\tanh(\alpha/2)}
\end{equation}
where $i$ is the grid index, $N$ is the total number of the nodal points. The factor $\alpha$ should be calculated using the first grid spacing at a starting point ($i=1$), and this spacing is set sufficiently small to capture the refined structure near apparent contact line. The $[0,\pi]$ domain is split into two segments with a common starting point. The location of the starting point should be specified to implement the distribute function (\ref{eq:disfun}) for each segment. It can be selected using the point which has the maximum rate of change in the whole free-surface profile. A function like the denominator on the right side of (\ref{eq:adpfun}) can be used as a criterion to calculate the change rate at a given location
\begin{equation}\label{eq:chrate}
f(\theta)=\sqrt{1+\alpha h_{\theta}^{2}+\beta h_{\theta\theta}^{2}}
\end{equation}
The time step $\Delta t$ can be also adapted according to a formula
\begin{equation}\label{eq:adpfunt}
\Delta t=\frac{\Delta h_{\mathrm{max}}}{\mathrm{abs}(Q_{\theta})_{\mathrm{max}}}
\end{equation}
where $\Delta h_{\mathrm{max}}$ is a preset maximum permissible change of film thickness between each time step, $Q_{\theta}$ is the net flow flux calculated from an explicit scheme which uses the film thickness at current time step.

\bibliography{physflupaper}

\end{document}